\documentclass{amsart}
\usepackage{amsmath, amssymb, graphics, setspace,color,soul,graphicx}
\usepackage{appendix}
\usepackage{graphicx}
\usepackage{subcaption}
\usepackage{wrapfig}
\usepackage{xcolor}
\usepackage[margin=1in]{geometry}
\usepackage{booktabs}

\newcommand{\mathsym}[1]{{}}
\newcommand{\unicode}[1]{{}}

\newcommand{\Rey}{\mathcal R_e}
\newcommand{\A}{\mathcal A}
\newcommand{\partderiv}[2]{\ensuremath{\frac{\partial #1}{\partial #2}}}
\newcommand{\partderivtwo}[2]{\ensuremath{\frac{\partial^2 #1}{\partial #2^2}}}

\newcommand{\partderivmix}[3]{\ensuremath{\frac{\partial^2 #1}{\partial #2\partial #3}}}

\newcommand{\Er}{\mathcal E_r}
\newcommand{\bi}{\mathbf i} 
\newcommand{\bj}{\mathbf j} 
\newcommand{\bk}{\mathbf k} 
\newcommand{\bv}{\mathbf v}
\newcommand{\bx}{\mathbf x}
\newcommand{\bn}{\mathbf n} 
\newcommand{\rx}{\bar x}
\newcommand{\ry}{\bar y}\newcommand{\rz}{\bar z}
\newcommand{\ru}{\bar u}
\newcommand{\rv}{\bar v}

\newcommand{\rw}{\bar w}

\newcommand{\wof}{W_{\text{\begin{tiny}{{OF}}\end{tiny}}}}

\newtheorem{remark}{\bf Remark}[section]
\newtheorem{proposition}{\bf Proposition}[section]

\title{ %Chevron Flow Pattern in  an Active Nematic Liquid Crystal in Smectic-A Confinement. 
Chevron patterns in an active nematic liquid crystal film in contact with Smectic A }

\begin{document}

\author{M.Carme Calderer}
\address{Department of Mathematics, University of Minnesota, Minneapolis, MN 55455, USA}
\email{calde014@umn.edu}
\author{Lingxing Yao}
\address{Department of Mathematics, The University of Akron, Akron, OH 44325, USA}
\email{lyao@uakron.edu}
\author{Longhua Zhao}
\address{Department of Mathematics, Case Western Reserve University, Cleveland, OH 44106, USA}
\email{longhua.zhao@case.edu}
\author{Dmitry Golovaty}
\address{Department of Mathematics, The University of Akron, Akron, OH 44325, USA}
\email{dmitry@uakron.edu}
\author{Jordi Ign\'es-Mullol}
\address{Department of Materials Science and Physical Chemistry, Universitat de Barcelona, Barcelona, Spain}
\address{Institute of Nanoscience and Nanotechnology, IN2UB, Universitat de Barcelona, 08028 Barcelona, Spain}
\email{jignes@ub.edu}
\author{Francesc Sagu\'es}
\address{Department of Materials Science and Physical Chemistry, Universitat de Barcelona, Barcelona, Spain}
\address{Institute of Nanoscience and Nanotechnology, IN2UB, Universitat de Barcelona, 08028 Barcelona, Spain}
\email{f.sagues@ub.edu}

\begin{abstract}
We study a new mechanism of active matter confinement of a thin, active nematic sample consisting of microtubules, activated by Adenosine Triphosphate (ATP), placed between a  slab of passive  liquid crystal,  the compound 8CB, and water. The  8CB slab is kept at a temperature below the phase transition value between the nematic and the smectic A phases. The smectic A molecules are  horizontally aligned with an applied magnetic field, with their centers of mass arranged on equally spaced layers perpendicular to the  field. The contact with the active nematic prompts flow  in the smectic slab, along the direction parallel to the layers.   This flow direction is transferred back to the active nematic. We set up a model of such contact flow and make predictions on the experimentally observed pattern, from the point of view of asymptotic, linear and nonlinear analyses. We examine such results within the scope of the principle of minimum energy dissipation of the flow. For analytic convenience, we consider the active nematic confined between two symmetric  8CB  slabs, and show that the conclusions  still hold when replacing the bottom smectic A substrate with water, as in the experimental setting.

\end{abstract}

\maketitle

\section{Introduction}
\begin{figure}[htb!]
    \centering
    \includegraphics[width=0.8\textwidth]{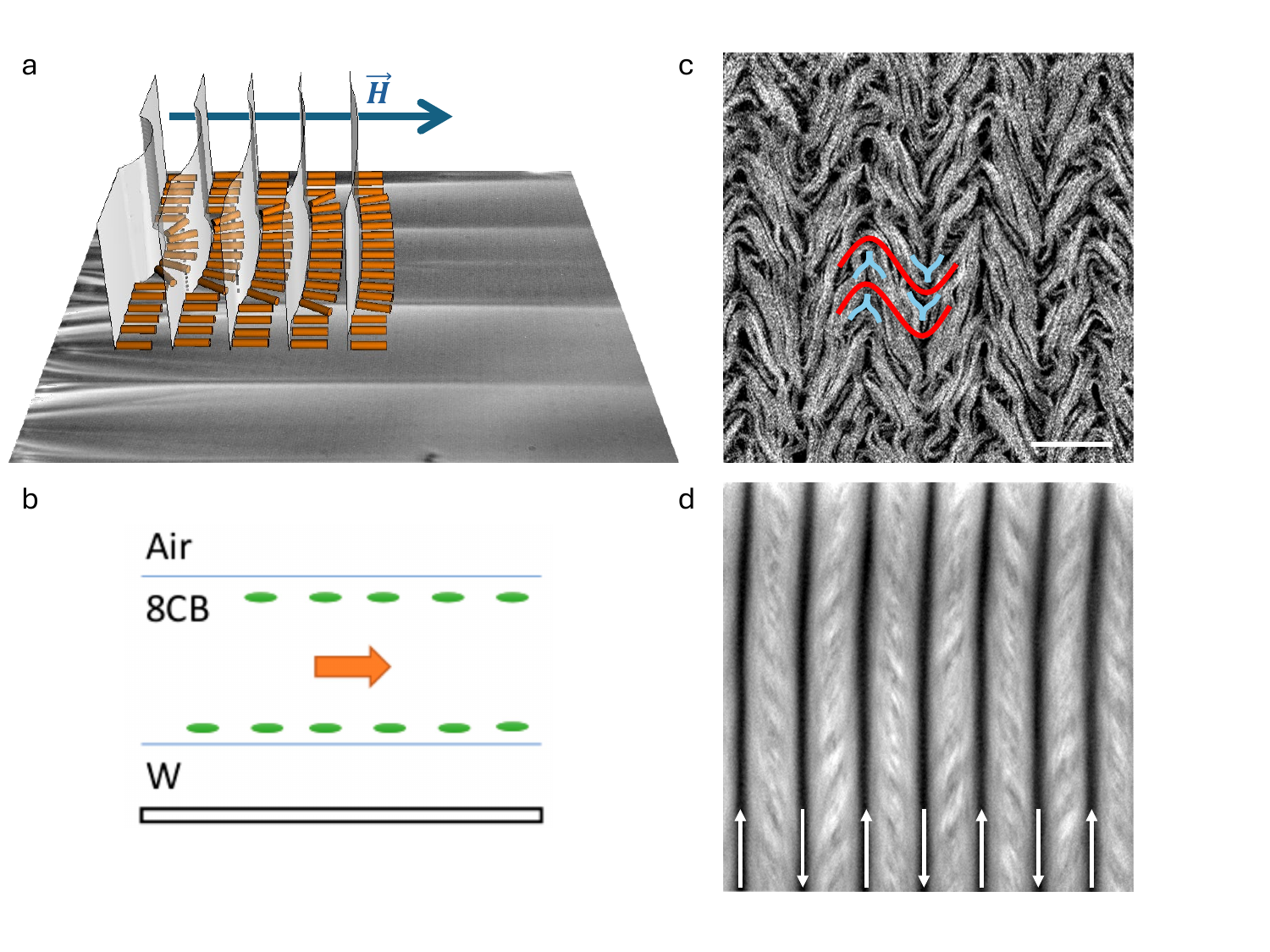}
     \caption{  (a) Polarizing optical micrograph, and configuration of the underlying molecular planes in the Sm A phase of the passive liquid crystal.   (b) A side view of the experimental domain. The orange arrow represents the direction of the applied magnetic  field, also the direction of alignment of the Smectic A molecules (8CB, green), in contact with air (top) and with water (W, bottom). The active nematic liquid crystal is confined between the Smectic A  and the water layers. 
       (c) Fluorescence micrograph of a close up of the $\pm \frac{1}{2}$ defects traveling along the ridges of the chevron pattern of the active nematic liquid crystal (Scale bars: 100 $\mu m$). (d) Time average of the dynamic pattern. The arrows illustrate the anti-parallel flow directions along the lanes of defect cores.
     }
      \label{fig1}
    \end{figure}

We study a new mechanism of active matter confinement by which an active nematic sample, consisting of microtubules, is placed between two thin slabs of passive liquid crystal,  the  8CB compound. The confining slabs of 8CB are kept at a temperature below the phase transition threshold between the nematic and the smectic A phases with its  molecules  horizontally aligned with an applied magnetic field. The family of smectic A layers, arranged perpendicularly to the magnetic field, and with an interspace  distance of about 30.5 A,  determine the flow direction of the liquid crystal. Specifically, the high viscosity of a smectic A liquid crystal for flow with velocity field perpendicular to the layers, effectively hinders the motion in such a direction, restricting it to the direction parallel to the layers. With no flow inducing mechanism, the smectic liquid crystal would be at rest. However, the contact with the active nematic induces flow in the smectic slabs, along the allowed direction, that is, parallel to the smectic  layers.  This flow direction is transferred to the active nematic  at the  contact interface. There is an active feedback between the smectic and the nematic liquid crystals: whereas the smectic layer, powered by a magnetic field aligns the nematic prompting it to flow, the latter, in turn maintains the smectic flow that preserves the nematic alignment. 
The work is based on the laboratory experiments and theoretical findings reported in   \cite{guillamat2017control, guillamat2016}, shown in Figure\ref{fig1}  within the context of turbulence control in active nematic by soft interfaces \cite{guillamat2017taming}. 
    
In this work, we analyze the governing equations of liquid crystal flow to make predictions on the experimentally observed chevron like structures. 
The latter consist of a family of   stripes, {\it chevrons},  on horizontal planes parallel to the substrates, aligned on the direction perpendicular to the applied magnetic field. The flow takes place on the direction parallel to the stripes and its direction alternates across neighboring bands. The chevron phenomenon occurs within a range of the activity parameter, provided that the magnetic field aligning the smectic A phase is preserved.  Values of the activity parameter above a  certain threshold lead to distortion of the bands with  an eventual  transition to turbulence \cite{guillamat2017taming}. 
In the laboratory, the active nematic/passive liquid crystal system  was prepared in a cylindrical pool of diameter 5 mm and depth 8 mm. The experiments revealed a scaling law of the velocity as $v\approx \alpha$, where $\alpha= \ln([\text{ATP}])$ measures the activity.  (The scaling law in the turbulent regime is $v\approx\sqrt {\alpha}$). 

Although defects are ubiquitous  in active liquid crystals, especially in their connection to turbulence, the confinement in sufficiently thin domains gives rise to laminar flow with  singularities being mostly non-prominent. However, the experiments clearly show that on the higher activity range but with the chevron structure still preserved, motion of defects of alternating degree $\pm \frac{1}{2}$ along parallel lines takes place. The work presented here neglects point defects, allowing us to model the flow according to the Ericksen-Leslie equations.  

A  mathematical model of the observed phenomenon involves the governing equations of active nematic in the region $\Omega_N= \{(x, y, z):  0< x^2+z^2<L^2,  -\frac{d}{2}<y<\frac{d}{2}\}$, $L\gg d>0,$  as in the experimental set up ($d$ is estimated to be between  5 to 10 $\mu$m). We let $\bi, \bj $ and $\bk$ denote the set of Euclidean orthonormal vectors.  The governing equations of the smectic liquid crystal are satisfied in the regions $\Omega_A= \{(x, y, z):  0< x^2+z^2<L^2,\,  y\in (\frac{d}{2}, \frac{d}{2}+d_0) \cup (-\frac{d}{2}-d_0, -\frac{d}{2})\}$,  where $d_0\approx 6$ mm denotes the thickness of the smectic region. Boundary conditions are prescribed at the interface $y=\pm \frac{d}{2}$.  %The smectic A region is in contact with air on top and with a layer of water and surfactant at the bottom layer.
An assumption of the model supported by the laboratory experiments is that  the two liquid crystals do not mix and keep a very well defined horizontal contact interface. 
The balance of linear momentum about these interfaces establishes the 
continuity of velocity fields, $\bv_N=\bv_A$, with the components perpendicular to the interface being zero, and the continuity of the forces. Since the interface is not solid, we impose a weak boundary condition on the nematic director field, that is, $\bn\cdot \bj=0$,  expressing tangentiality to the interface.  
The applied magnetic field maintains the direction of the smectic vector field $\bn_A$ along the $\bi$-direction, suppressing the flow on such direction (indeed, the Misowicz viscosity $\eta_1$ corresponding to the horizontal flow is about 150 times $\eta_2$ and $\eta_3$ corresponding to the perpendicular flow). Furthermore, applying the scaling property $d\ll d_0$ to simplify the equations in the nematic region,  under appropriate assumptions on the viscosity coefficients,  we arrive at the existence of a unique solution of the form $\bn_N=\bn(x,y)=(\cos\varphi(x,y),0, \sin\varphi(x,y)), \, \bv_N= \bv(x,y)= (0, 0, w_N(x))$ in the nematic region $\Omega_N$ and $\bv_A=(0, 0, w_A(x)$ in $\Omega_A$.

For both the nematic and smectic flow, the viscosity coefficients are required to satisfy Leslie's inequalities that guarantee the positivenss  of the rate of dissipation function of  the flow. However, such inequalities alone do not ensure existence of solutions for the smectic A flow, that require the positiveness of the Miesowicz viscosities. These turn out to be necessary and sufficient for the ellipticity of the governing system. In fact, the hydrodynamic system governing the smectic A flow is an anisotropic version of the Stokes problem of Newtonian fluids. To our knowledge, this is a new finding. 

The parametrization of the nematic field in terms of the angle of  alignment $\varphi(x)$  further reduces the governing system of the nematic to 
a second order of ordinary differential equation with oscillatory properties of solutions analogous to the Sturm-Liouville systems that model shear flow regimes of aligning passive nematic. The wave number of the solution is proportional to the activity parameter, in agreement with the experimental results. Moreover, we find that the velocity field is also oscillatory and periodically changes sign. 
The  equilibrium solutions  $\varphi=0,  \frac{\pi}{2},\,
0<\varphi=\varphi_0<\frac{\pi}{2}$ and $\varphi_1=\pi-\varphi_0$ play relevant roles in the analysis.  

Two main dimensionless parameter groups are identified as characterizing the flow, the Ericksen number $\Er$ (the ratio between the viscous and elastic torque) and the activity number $\A$, representing the strength of the ATP concentration driving the system. Comparison between the model and the experiments provides estimates on the numerical values of such parameters. Specifically, we find that $\A\Er\approx 10^4$ and $\Er\approx 10^2$, the former characterizing the strength of the angular momentum and $\Er$ that of the linear momentum. 
A key difference emerges with polymeric liquid crystals that may also form chevron structures under shear flow: the $\Er$ number of the latter can be of the order of $10^9$, reflecting the role of the viscous torque in driving the phenomenon. 
%%% Limiting regimes; principle of minimum dissipation.

In order to fully characterize the chevron, we  focus on the interface separating two neighboring bands. For this, we start observing that 
the governing equation for the angle of director alignment $\varphi$ is singular with respect to the parameter $\A\Er$, which estimates  the thickness value of the interface separating two neighbouring chevron bands to be approximately $\sqrt{\A\Er}\approx 10^{-2}$. From a related point of view, the asymptotic analysis at the limit 
$\A\Er\to 0$, yields the customary exponentially small internal layer transition. A subsequent analysis of the rate of dissipation function, when the flow (angle and velocity field) is approximated by the leading term of the asymptotic expansion, fails to select the pattern wave length. Specifically, the rate of dissipation function turns out to be  increasing on the wave length and it gets minimized at zero, that is, when no pattern is present. We also estimate the width of the interface by means of the heteroclinic orbit connecting two consecutive  angular domains, yielding a much larger value than that predicted by asymptotic analysis.  In a third approach, included in the Appendix, 
we postulate the separating interface as a (plane) defect, with a jump on the rate of dissipation function proportional to that of the velocity. 
Such an approach is borrowed from two different sources, involving an energy jump, one in static line defects  in liquid crystals and the second from phase transformations in solid mechanics \cite{ericksen1992introduction}. 

Let us now comment on the rate of dissipation analysis. This follows Rayleigh's theorem that characterizes a Newtonian steady state shear flow as that  minimizing the total rate of dissipation. Following the later generalization by Korteweg to rate type flow with internal variables, an idea that  became ubiquitous in the flow literature, we apply it to the study the chevron flow, and use it as another mechanism to select the wave length of the pattern.  For instance,  this approach was previously applied to the selection of   the wavelength of undulations of two-dimensional flows with suction of mass within  Darcy's  and  Stokes' frameworks \cite{ben2009suction}.  

In studies of active as well as passive liquid crystals, one main difficulty is the characterization of the Leslie coefficients, with the rotational coefficient $\alpha_1$ being particularly challenging. In the present case, such a coefficient is associated with a symmetry breaking property that assigns different flow speeds on channels with opposite flow direction. Since such a speed gap, even if existing, it is too small to be measured, it allows us to take $\alpha_1=0$.

%Arguments of data fitting are ubiquitous throughout this work. 

This paper is organized as follows. In section 2, we describe the model, including  basic properties of the smectic A liquid crystal flow.  In section 3, we develop the governing equations in the experimental geometry, their dimensionless setting and identify the main parameter groups of the system. In section 4, we study the chevron geometry, arriving at the  system of ordinary differential equations (ODEs) that models the angle of molecular alignment of the chevron. Three analytic approaches are developed, the linear oscillation, the singular perturbation internal layer structure and the nonlinear oscillatory pattern. 
In section 5, we present an example of the optimization of the total rate of energy dissipation  upon approximating it by the results of the linear analysis. Conclusions are left to section 6. An Appendix with  separate calculations is also included. 

Taking advantage of the  symmetry, we develop our analysis  in the case that the active nematic is surrounded by two smectic A layers. However, replacing the bottom one by water as in the actual experimental conditions, yields the same conclusions.  

\section{Acknowledgements}
M.C.C. acknowledges support from NSF grant DMS-1816740. J.I.-M., and F.S. acknowledge funding from MICINN/AEI/10.13039/501100011033 (Grant No. PID2022-137713NB-C21) and D.G. acknowledges funding from DMS-21065551. 

\section {A Model}  Previous to formulating the equations of the governing system, we discuss the Ericksen-Leslie equations of nematic flow and the equations of smectic A liquid crystal flow.  Both systems have the same structure, Lagrangian dissipative  systems, but differ in the form of the viscous component of the Cauchy stress tensor, and, in particular, in the number and values of the viscosity coefficients. 
In addition, the equations of smectic A also include that of the complex order parameter $\zeta$ that describes the geometry of the layers. However, the role of such an equation in the present work gets superseded by the fact that the layers are kept fixed by the applied magnetic field. 

\subsection{Nematic liquid crystal flow} 
\begin{figure}[htbp!]
\begin{subfigure}[b]{0.45\textwidth}
\includegraphics[width=1\textwidth]{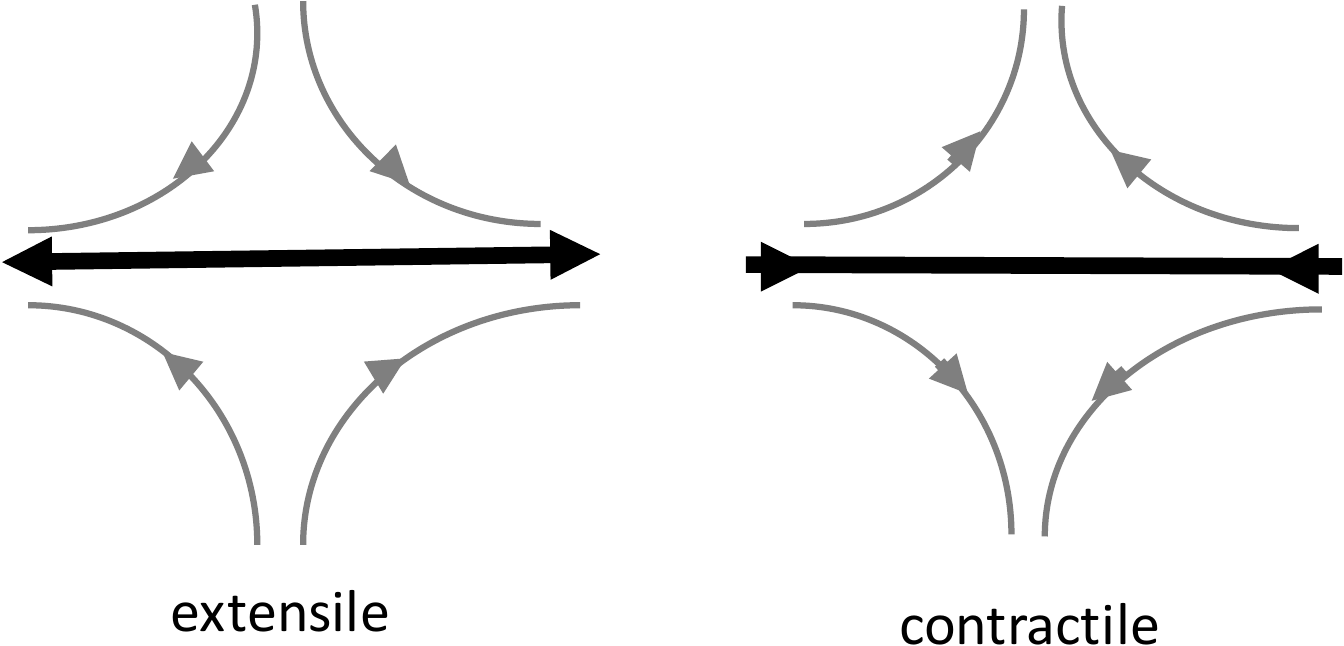}
\caption{fibers}
\label{push-pull}
\end{subfigure}
\hfill
\begin{subfigure}[b]{0.4\textwidth}
\begin{center}
    \includegraphics[width=0.6\textwidth]{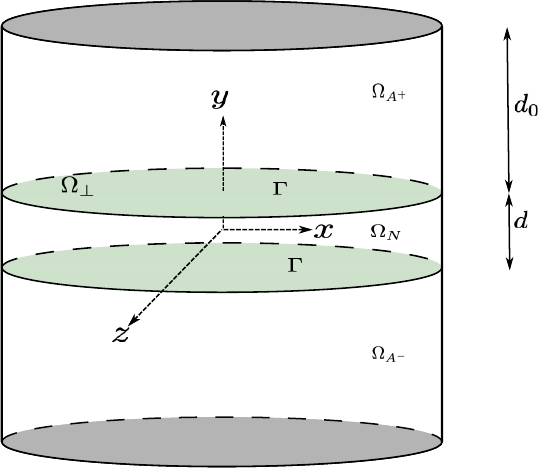}
    \caption{Sketch of the model domain}
    \label{domain-cylinder}
    \end{center}
    \end{subfigure}
    \caption{(\ref{push-pull}) Flow profile (grey curves) generated by  extensile (left) and contractile fibers (right). The thick black arrow represents the nematic director, pointing along the rod axis in rodlike particles, and  perpendicular to the disk in discotic ones.
    (\ref{domain-cylinder}) Sketch of the  domain $\Omega=\Omega_{A^+}\cup \Omega_N\cup \Omega_{A^-}$. The subdomain $\Omega_N$ represents the active nematic domain and $\Omega_{A+} $ and $
\Omega_{A^-} $ the upper and lower smectic domains, respectively,  with $\Gamma$ denoting the circular physical  interface between the nematic and the smectic A.  $\Omega_\perp$ denotes the cylindrical cross-section.}
\end{figure}

Similar to its passive counterpart, an active liquid crystal is assumed to be a viscous anisotropic and incompressible fluid with activity sources drawn from internal mechanisms or from the environment.
Let $\Omega\subset {\mathbb R}^3$ be an open domain occupied by the liquid crystal with boundary $\partial\Omega$. 
The Ericksen-Leslie equations of balance of linear and angular momentum, and the incompressibility and unit director constraints for the velocity  field $\bv$, pressure $p$ and  director  field $\bn$ in $\Omega$ and at time $t>0$ are \cite{leslie1992continuum,Calderer1996, Calderer1997}:
\begin{eqnarray}
\rho\dot\bv&=&\nabla \cdot \sigma_N, \label{lin-momentum}\\
\gamma_1 \dot{\mathbf{n}}\times \mathbf{n}&=&\nabla \cdot  \left(\frac{\partial  \wof}{\partial  \nabla \mathbf{n}} \right)\times \mathbf{ n}-\frac{\partial
 \wof}{\partial  \mathbf{n}}\times \mathbf{n}
 %\nonumber\\&&
 +\gamma_1 \boldsymbol{\Omega}  \mathbf{n}\times \mathbf{n} -\gamma_2 \mathbf{An}\times \mathbf{n},\label{ang-momentum}\\
\nabla\cdot\bv&=&0, \quad  %\label{incompressibility}\\
\bn\cdot\bn=1, \label{unit-length}
\end{eqnarray}
with $\rho>0$, constant, denoting the mass density, and $\gamma_1$ and $\gamma_2$ being combinations of Leslie coefficients defined in \eqref{alpha}. 
We point out that, since the system is strongly dissipative,  rotational inertia has been neglected in equation  (\ref{ang-momentum}). Moreover, such an equation results from taking the cross product of the  original equation of balance of angular momentum by the vector $\bn$. This has the advantage of explicitly suppressing the Lagrange multiplier associated with the unit director field constraint. 
 The function $\wof$ denotes the  Oseen-Frank energy of the liquid crystal, quadratic in the gradients of $\bn$:
 \begin{eqnarray}
\wof(\bn, \nabla\bn)&=& \frac{1}{2}\big(k_{11}|\nabla\cdot\bn|^2+ k_{22}|(\nabla\times\bn)\cdot\bn|^2
 \label{Oseen_Frank}
 %\\&&
 + k_{33}|(\nabla\times\bn)\times\bn|^2 \nonumber\\ &+&  (k_{22}+k_{24}) \nabla\cdot[(\bv\cdot\nabla)\bn-(\nabla\cdot\bn)\bn]\big), %\nonumber
 \end{eqnarray}
with $k_{11}$, $k_{22}$, $k_{33}>0$, $k_{22}>|k_{24}|$ and $2k_1\geq k_{22}+k_{24}$  denoting the Frank elastic constants.  The total energy of the system 
is $$\mathcal E=\int_{\Omega}\left(\frac{1}{2}\rho  \bv\cdot \bv+ \wof(\bn,\nabla\bn)\right)\,d\mathbf x.$$ 
 The Cauchy stress tensor  $\sigma_N$ is the sum of the elastic, viscous $\hat \sigma_N$ and active $\sigma_a$ components, respectively,  
\begin{eqnarray}
 \sigma_N&=&-p \mathbf{I} -\nabla \mathbf{n}^T\frac{\partial \wof}{\partial \nabla \mathbf{n}} 
+\hat{\sigma}_N+\sigma_a,\label{stress-Cauchy-total}\\
 \hat{\sigma}_N&=&\alpha_1\left(\mathbf{n}\cdot \mathbf{A}\mathbf{n}\right)\mathbf{n}\otimes  \mathbf{n}+ \alpha_2\mathbf{N}\otimes\mathbf{n}+\alpha_3\mathbf{n}\otimes \mathbf{N}+\alpha_4 \mathbf{A}%\nonumber\\&&
+\alpha_5 \mathbf{A} \mathbf{n}\otimes \mathbf{n}+\alpha_6\mathbf{n}\otimes  \mathbf{A} \mathbf{n},\label{stress-Cauchy-viscous}\\
\sigma_a&=&-a\,  \mathbf{n}\otimes \mathbf{n}, \label{stress-active}
\end{eqnarray}
where  $$2\mathbf{A} =\nabla \mathbf{v}+(\nabla \mathbf{v})^T,
\quad  2\boldsymbol{\Omega}=\nabla \mathbf{v} -(\nabla \mathbf{v})^T \mbox{ and }
\mathbf{N}=\dot{\mathbf{n}}-\boldsymbol{\Omega}  \mathbf{n}.$$
Here the superimposed  dot denotes the material time derivative, that is, $\dot f(t, \bx)= \partderiv{f}{t} + (\bv\cdot\nabla) f.$ 
 The Leslie coefficients $\alpha_i$, $1\leq i\leq 6$ represent the anisotropic viscosities of the liquid crystal. In particular, $\alpha_4$ corresponds to the isotropic or Newtonian viscosity, and 
 \begin{equation}
\gamma_1=\alpha_3-\alpha_2, \quad \gamma_2= \alpha_6-\alpha_5, \label{viscosity-rotational}
 \end{equation}
 denote the rotational (or twist) viscosity and the torsion coefficient, respectively. 
 The parameter $a$ in (\ref{stress-active}) quantifies the activity of the system,  
 with $a=0$ corresponding to the standard Ericksen-Leslie system for passive liquid crystals.
 The active part (\ref{stress-active}) of the stress tensor accounts for the dipolar non-conservative forces generated by the individual fibers. Their expressions reflect the symmetry of the flow field that they promote, with $a>0$ corresponding to the extensile regime, and $a<0$ to the contractile one \cite{Ramaswamy2010,Yeomans2007PhRvL,Marchetti2013, Edwards_2009} as illustrated in Fig \ref{push-pull}.   In the terminology of {\it swimmers}, extensile particles are known as {\it pushers} and contractile ones as {\it pullers}.  Note that $a$ has the dimensions of stress and is proportional to the ATP  concentration. Let us set the dependence of the stress activity  parameter $a$ on the ATP concentration as 
 \begin{equation}
     a:= c\,\mathcal C_0,  \quad \alpha=\log c\label{activity-concentration}
 \end{equation}
where $[c]=\frac{\text{mols of ATP}}{m^3}$ with  $\mu$M units, and $\alpha$ is known as the chemical activity. The scalar $\mathcal C_0$ is a material dependent function, with dimensions $[\mathcal C_0]= \text{stress}/\mu M$=$ \text{energy}/\text{mol}$, representing the energy (in kiloJoules) that 1 mol of ATP conveys to the system. It is also expected to depend on other parameters of the material.

%{\bf {ellipticipativeness of the system.}}

 The second law of thermodynamics in the form of the Clausius-Duhem inequality yields the entropy production inequality 
 \begin{equation}
     \hat\sigma_{N_{qp}}A_{pq} -(\gamma_1\mathbf N+\gamma_2 \mathbf A\bn)\cdot \mathbf N\geq 0. \label{entropy-production-ineq}
 \end{equation}
Leslie \cite{leslie1992continuum} has shown that this relation is satisfied provided the following inequalities hold:
\begin{eqnarray} \label{alpha}
\left\{
\begin{array}{r}
\alpha_4>0,\\
\alpha_1+\frac{3}{2}\alpha_4+\alpha_5+\alpha_6>0,\\
2\alpha_4+\alpha_5+\alpha_6>0,\\
4\gamma_1(2\alpha_4 +\alpha_5+\alpha_6)\geq (\alpha_2+\alpha_3-\alpha_6+\alpha_5)^2,\\
\gamma_1>0. %:=\alpha_3-\alpha_2>0.% \\
%\gamma_2:=\alpha_6-\alpha_5.
\end{array}
\right. \label{alpha-inequalities}
\end{eqnarray} 
 An immediate consequence of the latter is the positiveness of the  rate of dissipation function,  a quadratic  expression on the time-rate quantities ${\bf N}$ and ${\bf A}$. It takes the form 
\begin{eqnarray}
\Delta_N&=&\frac{1}{2}\left( \alpha_1(\bn^\text{T}\mathbf{A}\bn)^2
   +\gamma_1|\mathbf{N}|^2+ (\alpha_5+\alpha_6)|\mathbf{A}\bn|^2+
 (\alpha_3+\alpha_2+\gamma_2)
 {\mathbf{N}}^\text{T}
 \mathbf{A}\bn+\alpha_4|\mathbf{A}|^2
 \right)\geq 0.
\label{Rel}
\end{eqnarray}
 Parodi's relation, a consequence of Onsager's reciprocal relations in the microscale description of liquid crystals, results in the additional relation 
\begin{eqnarray}
\alpha_6-\alpha_5=\alpha_2+\alpha_3. \label{Parodi}
\end{eqnarray}
(In this case, the fourth inequality in \eqref{alpha} is an immediate consequence of the third and fifth ones). 
This condition renders the rate of dissipation function  a potential of the viscous stress \eqref{stress-Cauchy-viscous}, that is,
$   \hat\sigma_N= \partderiv{\Delta_N}{\nabla\bv}.$
   In association with $\Delta_N$, we consider the total rate of dissipation 
   \begin{equation}
\mathcal R:=\int_{\Omega}\Delta_N(\bx)\,d\bx. \label{dissipation-total}   \end{equation}
In fluid mechanics, the function $\mathcal R$ is relevant within the context of the Helmholtz minimum dissipation theorem, stating that the 
steady Stokes flow motion of an incompressible fluid has the smallest rate of dissipation than any other incompressible motion with the same velocity on the boundary. A generalization to viscous flows of fluids with internal structure was carried out by Korteweg \cite{korteweg1883xvii} and Rayleigh \cite{rayleigh1913lxv}. In particular, it applies to the liquid crystal flow. 
   
We consider a class of liquid crystals able to align under the effect of flow of small velocity gradient. This 
requires that 
\begin{equation}
\left|\frac{\gamma_1}{\gamma_2}\right|=:\frac{1}{\lambda}\leq 1, \label{aligning}
\end{equation}
where $\lambda$ is known as the flow alignment parameter. It represents the ratio between the extensional and rotational effects of the shear flow, with the former dominating in the case $\lambda>1$ and so the director aligns along the flow direction. The tumbling regime corresponds to $\lambda<1$, with a prevailing rotational couple that prevents $\bn$ from choosing an aligning direction \cite{baalss1988viscosity-both,Edwards_2009}.

Due to conditions that will become clear later, we also require
\begin{equation}
    \gamma_2<0. \label {lambda2-negative}
\end{equation}
This, together with the last inequality in \eqref{alpha-inequalities} and \eqref{aligning}, and taking Parodi's condition \eqref{Parodi} into account, yields
\begin{equation}
\gamma_2-\gamma_1=2\alpha_2 <0. \label{alpha2-negative}
\end{equation}
It is well known that  measuring individual viscosity coefficients is not possible. In the 1930's  Miesowicz measured three groups of viscosities $\eta_1, \eta_2$ and $\eta_3$  corresponding to basic flow geometries:
(a) $\eta_1$ correponds to the flow when $\bn$ is parallel to $\bv$; (b) $\eta_2$ is the viscosity of the flow when 
$\bn $ is parallel to $\nabla\bv$, and (c) $\eta_3$  is measured in the case that $\bn$ is orthogonal to both $\bv$ and $ \nabla\bv$, where
\begin{align}
    \eta_1=\frac{1}{2}(\alpha_3 + \alpha_4 +\alpha_6),\quad 
    \eta_2=\frac{1}{2}(-\alpha_2+\alpha_4+\alpha_5), \quad \eta_3=\frac{1}{2}\alpha_4.
\end{align}

As in \cite{calderer2021shear}, we choose the following values of the Leslie coefficients (in units of $\eta_0$, the typical viscosity) for the active nematic:
   \begin{eqnarray}
      && \alpha_1=0,\, \alpha_2=-1.5, \, \alpha_3=-0.5, \,\alpha_4=2,\nonumber\\
     && \alpha_5=2, \, \alpha_6=0, \, \gamma_1=1, \, \gamma_2=-2. \label{viscosity-values}
   \end{eqnarray}
   (Such scaled viscosities will be denoted with a super-imposed bar). % {\color{blue} These values have been updated later in the draft}.
  % \noindent {\color{red}{ Note.\,  The  above coefficients   satisfy the required Leslie inequalities and Parodi conditions. However, it may be relevant to change their scales. Ideally, $\alpha_4\sim \eta$, where $\eta$ is the typical viscosity, which is well known for the fiber flow under study. For the 8CB, the authors take  $\eta= 30$mPa.s \cite{guillamat2017taming}.  On the other hand, the Leslie coefficients for 8CB in both, the nematic and smectic phases, have been measured. It may be wise to use those values instead of the simplified numbers listed above.  We should be cautious about it, though. In the experiment, 8CB is layered on top of the nematic filaments, and so, the 8CB viscosity is different than that of the filaments. }}
\subsection {Flow of Smectic A liquid crystals} These are liquid crystals with molecular centers of mass placed in equally spaced one-dimensional layers and may flow along them,  with the directions of the molecules  aligned perpendicularly to such layers. That is, smectic A liquid crystals  behave like a solid crystal in one direction and liquid in the two perpendicular ones. Effectively, this behavior is the result of  the remarkable anisotropy properties of the viscosity, with the coefficient corresponding to the  flow along the direction perpendicular to the layers being much larger than that in the other two directions, and in fact, diverging to infinity.

The previously stated  facts on smectic A flow can be rigorously interpreted in terms of the Miesowicz viscosity measurements, by the observation that $\eta_1$ measures the flow viscosity  on the direction perpendicular to the layers and $\eta_2$ that of  the flow parallel to the layers. 
Measurements studying  the dependence of the  viscosities with respect to the  temperature followed the work of Miesowicz. In particular, at the temperature of  transition to the smectic A  phase,  it was found that  $\frac{\eta_1}{\eta_2}= 150$,  (\cite{leger1976viscosity}, figure 7) confirming the divergence of $\eta_1$, which prevents flow perpendicular to the layers.  

To model equilibrium configurations of smectic A liquid crystals, we follow  the theory proposed by de Gennes, that in addition to the director field $\bn$  incorporates the complex order parameter $\psi$,   previously postulated by Mcmillan \cite{de1993physics}, \cite{mcmillan1971simple}.
%(A generalization of the later to include nonlinear terms  and an analysis of the phase transitions can be found in \cite{biscari2007landau} and \cite{chen1976landau}).  
Equilibrium configurations of a smectic A liquid crystal, occupying a domain $\Omega\subset {\mathbb R}^3$, are described by pairs of fields $(\bn, \psi:=\rho_0 e^{i\zeta})$, with $\rho_0 $ and $\zeta$ representing the modulus and the phase of the complex order parameter, respectively. 
The positive quantity $\rho_0$ gives the density of molecules in the smectic phase and the level surfaces $\zeta=\text {constant}$ represent the smectic layers.
In the undistorted equilibrium state $\nabla\zeta=q \bn$ holds, with $\rho_0=1$ encoding the geometric signature  of the smectic layers, with an inter space separation $\frac{2\pi}{q}$, and $\bn $ locally perpendicular to the layers. An external magnetic field $\mathbf M$ applied perpendicular to the layers contributes to further  align the molecules along the field direction,  maintaining the layer structure. Indeed, the contribution of the magnetic field to the energy (per unit volume) is $-\chi\mathbf M\cdot \bn$, with $\chi>0$ representing the magnetic permeability of the material. The  application of the magnetic field is pivotal to the experiment that we study, since failure of such a field would suppress the chevron structure. 

The flow theory of Leslie postulates $\{\bn, \mathbf a:=\frac{\nabla\zeta}{|\nabla\zeta|}, \bv, p\}$ as the fields to characterize a smectic A flow configuration \cite{leslie1999theory}, \cite{stewart2007smectic}. As in the nematic case, $\bn, \bv$ and the scalar $p$ represent the director and velocity vector fields, and the scalar $p$  the pressure, respectively. 
The unit vector $\mathbf a$, introduced by de Gennes  \cite{de1993physics}, is  locally perpendicular to the layers. Introducing it as an independent field rather than through the constraint relation $\mathbf a= q \bn$ (with $|\nabla\zeta| :=q$) allows for breaking the smectic A constraint in studies of phase transitions  \cite{biscari2007landau}, \cite{calderer2008continuum} to the nematic or to the C-phases.

In this work, we apply the model of flow of smectic A liquid crystals as in \cite{stewart2007smectic}. The governing equations are as in (\ref{lin-momentum})-(\ref{unit-length}) but with  $\sigma_N$ replaced with $\sigma_A$.  (These equations correspond to (4.13)-(4.21) in \cite{stewart2007smectic}).  We present the expression of $\sigma_A$ in the Appendix. 
Rather than explicitly showing the governing equations in their general form, we will make use of the significant simplifications due to the special bookshelf geometry, maintained by the applied magnetic field.  Specifically,  we are interested in solutions such that the layer normal vector $\nabla\zeta=q\,\bi$. In particular, this reduces the stress tensor $\sigma_A$ to having  the same form as $\sigma_N$
 but with different viscosity coefficients $\alpha_i,$ $i=1,\dots 6.$ 
 
 We also note that 
looking for solutions with constant director and layer normal makes the additional equation for the complex order parameter trivial. 

We now set up the governing equations for the  fields
\begin{align} \label{smectic-stationary-flow}
\bn=\mathbf i, \quad \mathbf a=\bi,  \quad \bv=(0, v(x, y,z), w(x, y, z)), \quad   p=p(x,y, z).   
\end{align}
The stress tensor takes the form
\begin{equation}
   \sigma_A= -pI +\tilde\sigma_A\end{equation}
   which expressed in components becomes
   \begin{align} \label{stress-smecticA}
   2 \hat\sigma_A= &\alpha_2\left[\begin{matrix} 0&0&0\\-v_x&0&0\\-w_x&0&0\end{matrix}\right]+ \alpha_3\left[\begin{matrix} 0&-v_x& -w_x\\0&0&0\\0&0&0\end{matrix}\right] + \alpha_4\left[\begin{matrix} 0 &v_x&w_x\\v_x&2v_y &v_z+w_y\\ w_x & v_z+w_y &2 w_z\end{matrix}\right]\nonumber 
   +  \alpha_5\left[\begin{matrix} 0&0&0\\v_x &0&0 \\w_x&0&0\end{matrix}\right]
   \\&+\alpha_6 \left[\begin{matrix} 0 &v_x&w_x\\ 0&0&0\\0&0&0\end{matrix}\right]
   = \left[\begin{matrix} 0 & v_x(\alpha_2+\alpha_4+ \alpha_5) & w_x(\alpha_2+\alpha_4+ \alpha_5) \\ v_x(-\alpha_2+\alpha_4+\alpha_5) & 2\alpha_4 v_y& \alpha_4(v_z+ w_y)\\w_x(-\alpha_2+\alpha_4+\alpha_5) &  \alpha_4(v_z+ w_y)& 2\alpha_4 w_z
   \end{matrix}\right].
   \end{align}
In addition to requiring  the inequalities \eqref{alpha-inequalities},  we will also require that the viscosities of the latter satisfy 
\begin{equation}  \eta_2=\frac{1}{2} (-\alpha_2+\alpha_4+\alpha_5)> 0. \label{extra-condition-smectic}\end{equation}
In fact, this turns out to be the effective Miesowicz viscosity of the smectic A flow studied here. 

%In section 3, we will see  that for nematic flow such an inequality corresponds to  the dissipation relation \eqref{entropy-production-ineq} applied to  shear flow, in the the case that $\alpha_1=0$. 

\subsection{Interfacial conditions  between the smectic and nematic fluids} 
We start setting up the cylindrical domain $\Omega\subset\mathbb R^3$ of the two-phase liquid crystal system. The active nematic occupies the region $\Omega_N$ defined as follows:
\begin{align} \Omega_N=&\Omega_\perp \times \left(-\frac{d}{2}, \frac{d}{2}\right)=\{(x,y,z)\in\mathbb R^3: \,(x,z)\in\Omega_\perp, \, -\frac{d}{2}<y<\frac{d}{2}\}, \nonumber \\
\Omega_\perp&= \{(x,z): 0\leq  x^2+z^2\leq L^2\}. 
%(-L_x, L_x)\times (-L_z, L_z). 
\label{domains}\end{align}
As compared to figures \ref{fig1}(a) and \ref{fig1}(b), the $x$-axis is on the direction of the magnetic field (and that of the Smectic A molecules), the $z$-axis is parallel to the flow direction, and the $y$-axis is perpendicular to the Smectic A substrates.  
The region occupied by the Smectic A substrates is 
\begin{equation}
    \Omega_A:= {\Omega_{A^+}}\cup \Omega_{A^-}:= \{(x, z)\in \Omega_\perp, \, \frac{d}{2}<y<\frac{d}{2}+d_0\} \cup  \{(x, z)\in \Omega_\perp, \, -d_0-\frac{d}{2}<y<-\frac{d}{2}\}. \label{domainA}
\end{equation}
For  convenience, we will use the same notation $\Omega_A$ to denote either of the upper or the lower substrate.

Let  $\Gamma$ denote the interface between the nematic and the smectic liquid crystals, represented by the plane $y=\frac{d}{2}$ (likewise, $y=-\frac{d}{2}$ refers to the lower contact region, also called $\Gamma$). With such a representation, consistent with the domain selections \eqref{domains}-\eqref{domainA}, we implicitly assume that $\Gamma$ remains fixed. Furthermore, since no mass transfer is assumed to occur across $\Gamma$, a consequence of the immiscibility property between the fluids, then such properties imply the continuity of the velocity fields together with the vanishing of the components normal to $\Gamma$ \cite{batchelor1967introduction, leal2007advanced}:
\begin{align}
    & \bv_A=\bv_N , \label{continuity-velocity}\\
     & \bv_A\cdot \bj =0 =\bv_N\cdot \bj \quad \text {on} \quad \Gamma. \label{Interface-conditions}
\end{align}
The component form of these equations, taking \eqref{smectic-stationary-flow} into account, is
\begin{equation}
  u_A(x,\frac{d}{2}, z)=0=u_N(x,\frac{d}{2}, z), \quad   v_A(x, \frac{d}{2}, z)=0=v_N(x,\frac{d}{2},z), \quad w_A(x,\frac{d}{2},z)= w_N(x,\frac{d}{2},z).\label{cont-velocity-d}
\end{equation}

   In addition, continuity of the forces across the interface is also required,
   \begin{equation}
       T_A\, \mathbf j= T_N\, \mathbf j \quad \text{on} \quad \Gamma. \label{cont-forces}
   \end{equation}
   Next, we specify the boundary conditions on the top surface:
 \begin{enumerate}
     \item 
 At $y=\frac{d}{2}+d_0$, we assume that the smectic A is in contact with air at rest and sustaining a pressure $P_0>0$. We have
\begin{align}
        &\sigma_A\, \mathbf j =-P_0\,  \mathbf j, \label{force-balance-interface}\\
        &v_A(x, \frac{d}{2}+d_0, z)=0,  \quad w_A(x, \frac{d}{2}+d_0, z)=0.\label{zero-normal-velocity}
    \end{align}
\item It will be shown later that the vector  equation \eqref{cont-forces} reduces to the scalar one equating the pressures on both regions, that is
\begin{equation}
    p_N(x, \frac{d}{2}, z)= P_0=  p_A(x, \frac{d}{2}, z), \quad (x, \frac{d}{2}, z)\in \Gamma. \label{cont-pressure}
\end{equation}
    \end{enumerate}  
    We also assume a weak anchoring condition on the director field at the interface $\Gamma$, that is
\begin{equation}   
\bn_N(x, \pm \frac{d}{2}, z)\cdot\bj =0. 
\label{weak-anchoring}
\end{equation}
In particular, this yields 
\begin{equation}  n_2(x,\pm \frac{d}{2}, z) =0. 
\label{zero-n2-component}
\end{equation}

\section {Governing equations}

%There are two key points on the dynamics of this system: \begin{enumerate}
   % \item  The layering configuration of the smectic A liquid crystal prevents flow in the direction {\it parallel} to $\bn=\mathbf i$.  The experimental observation points to $\bv_A= v_A(x)\bk$.\item The smectic flow is powered by the active flow. In particular, $v_A=0$ when the activity is zero. \end{enumerate} The previous observations gives us guidance to study the smectic A flow.
\subsection{ Smectic A flow: dimensionless equations.}
  Appealing to the fact that $0<\Rey\ll 1$, we postulate the equations of the steady state flow of the smectic A.  We also start enforcing the director field  $\bn_A=\bi$, as sustained by the magnetic field, and also imposing the flow constraint of the smectic A in that the $x$-component of the velocity field satisfies $u=0$.
  Let us write down the equations of stationary smectic flow for the fields (\ref{smectic-stationary-flow}), in the domain $\Omega_A$:
    \begin{align} \label{governing-smectic-flow}
    0=& \partderiv{v_A}{y} + \partderiv{w_A}{z}, \nonumber \\
     %  \rho {v_A}_j\partderiv{{v_A}_i}{x_j}
       0= & -\partderiv{p}{x_i} + \partderiv{\hat\sigma_{ij}}{x_j}, \quad i=2, 3.
    \end{align}
    Note that, in the latter, we do not include the balance of angular momentum equation since we have already set $\bn=\bi$. Such an equation is identically satisfied upon the choice of the appropriate Lagrange multiplier. 

    The components of equation (\ref{force-balance-interface}), taking into account the relations  (\ref{stress-smecticA}) give 
    \begin{align}
      (\alpha_4+\alpha_5+\alpha_2)\partderiv{v_A}{x}(x, \frac{d}{2}+d_0, z)&=0, \label{eqn1}\\
        (-p+ \alpha_4 \partderiv{v_A}{y})(x, \frac{d}{2}+d_0, z)&=-P_0, \label{eqn2}\\
         \alpha_4(\partderiv{v_A}{z}+\partderiv{w_A}{y})(x, \frac{d}{2}+d_0, z)&=0, \quad (x,z)\in\Omega_\perp. \label{eqn3}
    \end{align}
    Note that (\ref{eqn1}) is a consequence of the first equation in (\ref{zero-normal-velocity}). The same equation in \eqref{zero-normal-velocity} also reduces relation  \eqref{eqn3} to
     \begin{equation}
         \partderiv{w_A}{y}(x, \frac{d}{2}+d_0, z)=0. \label{partderiv-y-topboundary}
     \end{equation}
     We see that  equation \eqref{eqn2} gives the value of $p$ on the top boundary.
    %\begin{equation}   (-p+\alpha_4\partderiv{v_A}{y})(x, d_0+d, z)=-P_0,  \quad (x, z)\in \Omega_\perp, \label{force-balance-interface-reduced}\end{equation} which can be regarded as assigning a value to $p$ at the top boundary.
   % The setting of the problem in the smectic A region is complete by specifying boundary conditions at the interface with the nematic. For convenience, we use the {\it N}-label to indicate fields associated with the nematic fluid.  So, the interface conditions at that contact become
  %  \begin{align} \label{Interface-nematic}    v_N(x, d, z)=&v_A(x, d, z) \nonumber\\ \sigma_N(x, d, z)\mathbf j= & \sigma(x, d, z)\bj.  \end{align}
Next, we express the equations (\ref{governing-smectic-flow}) in terms of their components. We also set the right hand sides equal to zero, reflecting the small value of the Reynolds number in the corresponding dimensionless equations:
    \begin{eqnarray} 
    0&=& -p_x + \frac{1}{2}(\alpha_2+\alpha_4+\alpha_5) \left(\frac{\partial^2v_A}{\partial x\partial y} + \frac{\partial^2 w_A}{\partial x\partial z}\right), \\
   0&=& -p_y +\frac{1}{2} (-\alpha_2+ \alpha_4 +\alpha_5) \partderivtwo{v_A}{x} +\alpha_4  \frac{\partial^2 v_A}{\partial y^2}+\frac{\alpha_4}{2} \left(\frac{\partial^2{v_A}}{\partial z^2} + \frac{\partial^2{w_A}}{\partial y \partial z}\right),  
   \\ 0&=&-p_z+ \frac{1}{2}(-\alpha_2 + \alpha_4+\alpha_5)\partderivtwo{w_A}{x} + \frac{\alpha_4}{2}\left(\partderivmix{v_A}{z}{y}+ \partderivtwo{w_A}{y}\right)+ \alpha_4\partderivtwo{w_A}{z}.
     \end{eqnarray}
    We now apply  the first equation in (\ref{governing-smectic-flow}) into the previous ones,  reducing the governing system of the smectic A flow, in the domain
    $\Omega_A$, to the following: 
    \begin{align}
    0=&\partderiv{v_A}{y}+ \partderiv{w_A}{z}, \label{governing-smectic-simplified1}\\
        0=&-p_x,\label{governing-smectic-simplified2} \\
       0 =&-p_y+ \frac{1}{2}(-\alpha_2+ \alpha_4 + \alpha_5) \partderivtwo{v_A}{x} + \frac{\alpha_4}{2}(\partderivtwo{v_A}{y}+\partderivtwo{v_A}{z}), \label{governing-smectic-simplified3}\\
   0= & -p_z +\frac{1}{2}(-\alpha_2 +\alpha_4 +\alpha_5) \partderivtwo{w_A}{x} +\frac{\alpha_4}{2}(\partderivtwo{w_A}{y}+ \partderivtwo{w_A}{z}). \label{governing-smectic-simplified}
    \end{align}
    We summarize the conditions that hold on the boundary of $\Omega_A$:
    \begin{enumerate}
        \item On the top boundary, equations \eqref{force-balance-interface} and \eqref{eqn1}-\eqref{eqn3} reduce to \eqref{partderiv-y-topboundary}, \eqref{eqn2} together with the original equations \eqref{eqn1}-\eqref{eqn3}.
        \item At the interface $\Gamma$, equations \eqref{cont-velocity-d} and \eqref{cont-pressure} hold. 
        \item The boundary condition $w_N(x, \frac{d}{2}, z)$ is still unknown. It will be determined upon solving the problem in the nematic region. 
    \end{enumerate}
   % \begin{align}&v_A=0=w_A \,\, \text {on} \,\, y=\frac{d}{2}+ d_0,\\&v_A=0\,\, \text {on} \,\, y=\frac{d}{2}\end{align}
We now observe that the assumptions $\eta_2, \eta_3=\frac{1}{2}\alpha_4>0$ guarantee that the  linear system  \eqref{governing-smectic-simplified1}-\eqref{governing-smectic-simplified}, together with the boundary conditions, is of elliptic type  as in the case of  the linear Stokes problem of Newtonian  flow. 
Appealing to the uniqueness of the solution of such a system, and taking into account that $v|_{y=\frac{d}{2}+ d_0}=0= v|_{y=\frac{d}{2}}$, we conclude that 
\begin{equation}
 v_A= 0, \quad \text {in}\,\, \Omega_A.
\end{equation}
    This together with equation  \eqref{governing-smectic-simplified1} gives
    $$w=w_A(x).$$
\noindent
In summary, the fields in the region $\Omega_A$ are of the form
\begin{equation} 
\bn_A=\bi, \quad \bv_A= w_A(x)\bk, \quad \psi= e^{iq x}, \quad (x,y,z)\in \Omega_A, \label{smectic-flow}
\end{equation}are solutions of the problem. Here   $q$ denotes the wave number of the smectic layers (recall that the interlayer separation in Smectic A is approximately 30.5 Angstroms, several orders of magnitude smaller than the chevron half distance).

\subsection{ Nematic flow: dimensional analysis and lubrication approximation}
The governing equations in the nematic slab $\Omega_N$  are as (\ref{lin-momentum})-(\ref{unit-length}) and are satisfied by the 
fields
\begin{equation}
    \bn_N=\bn(x,y,z), \,\, \bv_N= \bv(x, y, z), \, \, p=p(x,y,z), \label{nematic}
\end{equation}
$\bv_N=\{v_i\}_{i=1}^3=(u, v, w)$.  For convenience, from now on,  we drop the subindex `N' when referring to the components of the nematic fields.

Letting $d$, $d_0$ and L  be as defined in the previous section,  we denote $V$ the typical velocity, its value being that of the nematic along the channels. 
Other representative quantities involve $\eta:=\alpha_4$, a viscosity,  $K$ a typical nematic Frank constant, and $\rho$, the mass density.  Moreover, the activity parameter $a$ measures the concentration of the active ingredient ATP.  We define the dimensionless space and time variables as 
\begin{eqnarray}
\bar{x}=\frac{x}{L},\, \,  \bar{y}=\frac{y}{d}, \, \, \bar{z}=\frac{z}{L},\,
 \bar t=\frac{t}{L/V},  \,\, \varepsilon=\frac{d}{L}. \label{scaling1}
\end{eqnarray}
 The dimensionless velocity components and the pressure, respectively, are 
\begin{equation}\bar u=\frac{u}{V},\bar{v}=\frac{v}{V}, \, \, \bar w=\frac{w}{V}, \,\, \bar p=\frac{p}{P}, \label{scaling2}\end{equation}
where $P=\frac{V\eta}{L}$ is the typical choice of the pressure scale for a  viscosity dominated flow. 
The dimensionless  coefficients and dimensionless parameter groups are:
\begin{align}
\bar{\alpha }_i=&\frac{\alpha _i}{\rho L   V},\quad \bar{\beta }_j=\frac{\beta _j}{\rho L  V}, \quad
 \bar{\gamma }_j=\frac{\gamma _j}{\rho L  V},\quad \bar{k}_r=\frac{k_{r}}{\rho L^2  V^2}, \\
%\bar{\nu }=\frac{\nu }{\rho  V^2}=\frac{\mathcal{J} K}{\rho  L^2  {V}^2}, 
\mathcal A=&\frac{a L}{\eta  V},\quad \Er=\frac{ \eta L V}{K},\quad \mathcal {R}=\frac{ \rho L V}{\eta }, \label{scaling3}
\end{align}
$1\leq i\leq 6$, $1\leq j\leq 2$ and $r=1, 3.$
We introduce the dimensionless gradient operator as 
\begin{equation}
\bar \nabla=\left(\partderiv{}{\bar x}, \varepsilon^{-1} \partderiv{}{\bar y}, \partderiv{}{\bar z}\right)\label{nabla-bar}\end{equation}
Table \ref{experimental-values} lists representative parameters of the problem  obtained from  \cite{velez2024probing}  upon converting the 2-dimensional values given there to the three dimensional ones used in this work (estimating the thickness of the nematic layer to be 5$\mu$m). 
%\begin{enumerate}
 % {  \item Elasticity modulus: $K_{33}\approx K_{11}= K=10^{-10} \, \text{Newton}= 10^{-1} \text{gr}\,\mu{\text m}/{\text s}^2$
  %  \item Viscosity: $\eta= 0.4\text {Pa}*\text{s}= \frac{4}{3}10^{-2} \text{gr}/(\mu\text{m} \, {\text s}^2)$}
   % \item Apparatus diameter: $L= 5 \text { mm}= 5* 10^{3} \mu \text{m}$.
    %\item Velocity: $V=4 \mu m/\text{s}.$ This value corresponds to the maximum velocity measured in the experiment.
%\end{enumerate}

\begin{table}[h!]
\centering
%\begin{center}
\begin{tabular}{ |c|c|c|c|c|c|} 
 \hline\hline
 $\text{Quantity}$& Velocity& Viscosity&Elastic Modulus& Apparatus size &Ericksen No.\\
 \hline
Symbol&V& $\eta$& $k$ & $L$ &$\Er$  \\ 
Exp. Value &4&  $4/3* 10^{-2}$ & $10^{-1}$ & $5*10^{3}$ & $80$ \\ 
 Units& $\mu$m/sec& gr/($\mu$m $sec^2$) & gr\,$\mu$m/ $sec^2$& $\mu$m& 1\\ 
 \hline
\end{tabular}
\caption{We take equal Frank constants $k_{11}=k=k_{33}, k_{24}=0$ and $\eta=0.4 \text{Pa}\times \text{s}
.$  }\label{experimental-values}
\end{table} 
With the above data, we can estimate the  parameter as $\Er\approx 80$.
The analogous  quantity in  polymer liquid crystal flow is several orders of magnitude higher, well over $10^6.$

In summary, we can state the dimensionless equations for the scaled fields \eqref{scaling2} as the   original ones but replacing $\nabla$ with the scaled operator $\bar\nabla$ \eqref{nabla-bar}, and $\alpha_i$, and $k_i$  replaced by the corresponding quantities \eqref{scaling3} with a superimposed bar.  Furthermore, the density on the left hand side of \eqref{lin-momentum} is replaced with the Reynolds number $\mathcal R$ in the scaled equation, and the terms involving $\bar k_i$ are now multiplied by  ${\Er}^{-1}$. These equations are presented  in the {Appendix and labelled as 
\eqref{lin-momentum-scaled-new}-\eqref{stress-active-scaled-new}}. 

The lubrication approximation is formally obtained from the governing equations at the limit 
$\varepsilon \to 0$.  It immediately follows from {\eqref{incompressibility-scaled-new}} that 
\begin{equation}
    \partderiv{\bar v}{\bar y}=0.
\end{equation}
This combined with the boundary conditions at the interfaces $\Gamma$ with the smectic A, yield
\begin{eqnarray}
    \bar v(\bar x, \bar y, \bar z)=0, \quad  (\bar x, \bar y, \bar z)\in \bar \Omega_N. \label{0-velocity-y}
\end{eqnarray}

Next, taking limit as $\varepsilon\to 0$ in equation \eqref{ang-momentum1} and taking \eqref{scaled-Nforce-new} into account, we arrive at 
\begin{equation}
\partderivtwo{n_i}{\bar y}(\bar x, \bar y, \bar z)=0, \quad i=1, 2, 3. \label{eqns_n1n2n3}
\end{equation}
The equation for $i=1$, together with the zero boundary conditions on $n_2$ at the $\Gamma$-interface $y=\pm \frac{d}{2}$ ($\bar y=\frac{1}{2}$), give
\begin{equation}
    n_2(\bar x, \bar y, \bar z)=0, \quad (\bar x,\bar y,\bar z)\in \bar \Omega_N.
\end{equation}
We parameterize $\bn$ in terms of its angle of inclination $\varphi $ with respect to the horizontal,
\begin{equation}
    \bn=(\cos\varphi(\bar x,\bar y, \bar z), 0, \sin\varphi(\bar x, \bar y,\bar z)), \,\, \text {and let } \, \varphi_\perp:=\frac{\pi}{2}-\varphi\label{angular-representation}
\end{equation}
Applying these expressions into equations \eqref{eqns_n1n2n3} transform the latter into
\begin{equation}
    0=\sin\varphi\varphi_{\bar y\bar y}+\cos\varphi \varphi^2_{\bar y}, \quad  0=\cos\varphi\varphi_{\bar y\bar y}-\sin\varphi \varphi^2_{\bar y}.
\end{equation}
Multiplying the first equation by $\cos\varphi$ and the second one by $\sin\varphi$ and subtracting them, we get 
\begin{equation}\varphi_y(\bar x, \bar y,\bar z)=0, \quad  (\bar x, \bar y, \bar z)\in \bar \Omega_N. \end{equation}
Hence 
\begin{equation}
    \varphi(\bar x,\bar y, \bar z)=\varphi_0(\bar x, \bar z), \quad (\bar x,0, \bar z)\in \bar \Omega_N. \label{phi-2-indep-var}
\end{equation}
The asymptotic analysis of equations {\eqref{lin-momentum-scaled-new}} yield the leading equations 
\begin{equation}
    \partderivtwo{\bar u}{\bar y}=0= \partderivtwo{\bar w}{\bar y}. \label{leading-order-uw}
\end{equation}
Taking into account the first equation in \eqref{leading-order-uw} together with \eqref{smectic-stationary-flow} and the continuity of the  velocity across $ \Gamma$ \eqref{continuity-velocity}, gives 
\begin{equation}
    \bar u(x, y, z)=0 \quad \text{in} \quad \Omega_N. \label{0-velocity-x}
\end{equation}
Applying again {\eqref{incompressibility-scaled-new}}, taking \eqref{0-velocity-y}
 and \eqref{0-velocity-x} into account, we  arrive at 
 \begin{equation} \bar \bv_n=(0, 0, \bar w(\bar x, \bar y)). \end{equation}
Finally, we apply equation {\eqref{balance-lin-momenttum-leading3}} and require the solution to be symmetric respect to $y=0$, that is, requiring that $w|_{y=\frac{d}{2}}=w|_{y=-\frac{d}{2}}, $ in which case the integration of such  an equation gives us
\begin{equation}
    \partderiv{\bar w}{\bar y}=0 \quad \text {in} \,\, \bar \Omega_N, \label{y-symmetry}
\end{equation}
arriving at $\bar \bv=(0, 0, \bar w(x))$. 

\noindent {\bf Notational convention.\,} From now own, we suppress the upper bar notation on the dimensionless variables and quantities. Units will be included when referring to a dimensional quantity.  

In summary, the fields in the region $\Omega_N$ take the form
 \begin{equation}
     \bv_N(x, y, z)= w(x)\bk, \quad \bn_N(x, y, z)= (\cos\varphi(x), 0, \sin\varphi(x)), \quad p=p(x,z).    \label{nematic-reduced-form}
 \end{equation}
In section 4, we will solve the equations of the nematic to find $\varphi=\varphi(x)$, $
 w=w(x) $ and $p=p(x,z)$. The $z$-component $w(x)$ of the velocity field will then be applied as boundary condition in the second equation \eqref{cont-velocity-d}, allowing to complete the calculation of the fields in the smectic A region, and hence solve the entire problem in $\Omega$.

\begin{remark} \label{lower-water-substrate}
A straightforward calculation shows that the same outcome as in \eqref{nematic-reduced-form} follows in the case that the nematic layer is supported by water.  Assuming that non-slip boundary conditions hold on the bottom surface $y=-\frac{d}{2}-d_0$,  as well as requiring the continuity of the velocity fields and forces on $\Gamma^-$, the Stoke's problem 
in the region $(-\frac{d}{2}, -\frac{d}{2}-d_0)$ (labelled as $\Omega_A$ in figure 2 (b)) can be trivially solved to deliver the value $w_N|_{y=-\frac{d}{2}}.$ The possible quantitative difference between both approaches is resolved by taking the effective viscosity \eqref{g-function}  in the Stokes problem. 
\end{remark}

\section {Chevron Pattern} Let us now assume the nematic flow as in \eqref{nematic-reduced-form}, for which  the governing equations of balance of linear momentum take the form
  \begin{align}  
  \label{v1-eqn-reduced}
  0=& -\partderiv{p}{x} -2\Er^{-1}\partderiv{}{x}\left((\partderiv{n_1}{x})^2+ (\partderiv{n_3}{x})^2\right)\\ +& \, \, \frac{1}{2}\partderiv{}{x}\left(\partderiv{w}{x}n_1n_3(2\alpha_1 n_1^2+\alpha_2+\alpha_3 +\alpha_5+\alpha_6)-2{\mathcal{A}} n_1^2\right), \nonumber\\
     0=&-\varepsilon^{-1}\partderiv{p}{y}, \label{v2-eqn-reduced}\\
  0=& -\partderiv{p}{z}+\frac{1}{2}\partderiv{}{x}\left(\partderiv{w}{x}[2\alpha_1(n_1^2n_3^2+ n_1n_3^3)+ \alpha_2n_3^2-\alpha_3 n_1^2 + \alpha_4 + \alpha_5 n_3^2 +\alpha_6 n_1^2]-2{{\mathcal A}}n_1n_3\right). \label{v3-eqn-reduced} \end{align}
   The first and third components of the balance of angular momentum equations identically vanish, with the  second one formulated as 
   \begin{equation}      0= 2\Er^{-1}\left(\partderiv{}{x}(\partderiv{n_3}{x})n_1 - \partderiv{}{x}(\partderiv{n_1}{x})n_3\right)+\partderiv{w}{x}\left( \frac{\gamma_1}{2}- \frac{\gamma_2}{2}(n_1^2-n_3^2)\right). \label{n-eqn-reduce}\end{equation}
Differentiating \eqref{v1-eqn-reduced} with respect to $z$ gives
\begin{equation}   \frac{\partial^2{p}}{\partial x\partial z}=  0.  \end{equation}
 Likewise, differentiating the third equation \eqref{v3-eqn-reduced} with respect to $x$ and taking the former into account yields
  \begin{equation}  0=\frac{\partial^2}{\partial x^2}\left(\partderiv{w}{x}[2\alpha_1(n_1^2n_3^2+ n_1n_3^3)+ \alpha_2n_3^2-\alpha_3 n_1^2 + \alpha_4 + \alpha_5 n_3^2 +\alpha_6 n_1^2]-2{{\mathcal A}}n_1n_3\right). \label{general-ode}  \end{equation}
  Integrating the latter with respect to $x$ twice and setting the constants of integration equal to zero, we arrive at 
  \begin{equation}
    g(2\varphi)  \partderiv{w}{x}= 2{{\mathcal A}}  n_1 n_3  
      =
      \mathcal A\sin(2\varphi),\label{velocity_gradient-eqn-reduced}
  \end{equation}
  where the parametrization of $\bn$  has been applied in the last step as well as in the definition 
  \begin{align}
  g(2\varphi):= &{2\alpha_1}(\sin^2\varphi\cos^2\varphi{+}\sin^3\varphi\cos\varphi) + (\alpha_2+\alpha_5)\sin^2\varphi+ (\alpha_6 -\alpha_3)\cos^2\varphi + \alpha_4 \label{g-function} \\
  =&
\alpha_1\frac{1}{2}(\sin^2{2\varphi}{+}\sin{2\varphi}(1-\cos{2\varphi})) + \frac{1}{2}\big((\alpha_2+\alpha_5)(1-\cos{2\varphi})+ (\alpha_6-\alpha_3)(1+\cos{2\varphi})\big)+\alpha_4. \nonumber\end{align}
It follows from the entropy production inequality \eqref{entropy-production-ineq} that 
\begin{equation}
    g(2\varphi)\geq 0. \label{positivity-g}
\end{equation}

Furthermore, in the case that Parodi's relation holds, it reduces to
\begin{equation}
 g(2\varphi)= \alpha_1\frac{1}{2}(\sin^2{2\varphi}{+}\sin{2\varphi}(1-\cos{2\varphi})) + \alpha_2 + \alpha_5+\alpha_4,
\end{equation}
A simple calculation using Matlab shows that 
\begin{eqnarray}
& \alpha_2+\alpha_4+\alpha_5 -0.3612\alpha_1\leq g(2\varphi)\leq \alpha_2+\alpha_4+\alpha_5 +0.753\alpha_1, & \text {if}\quad \alpha_1\geq 0,\\
& \alpha_2+\alpha_4+\alpha_5 +1.0753\alpha_1\leq g(2\varphi)\leq \alpha_2+\alpha_4+\alpha_5 -0.3612\alpha_1, & \text {if}\quad \alpha_1\leq 0.
\end{eqnarray}

\begin{remark}
In the  forthcoming section on dissipation, we will justify the role of $g(2\varphi)$ as the effective viscosity of the flow. 
\end{remark}
We will further assume that $g$ is bounded  away from 0, that is 
\begin{equation}
    g(2\varphi)\geq \mu>0, \label{g-bounded-below}
\end{equation}
    where $\mu$ is some arbitrarily small positive number. For instance, an arbitrarily small increase of $\alpha_4$ with respect to its experimental value would provide such a lower bound on $g$. 
    
    The significance of choosing the integration constants identically zero in the integration of {\eqref{general-ode}} comes from the fact that the flow is not activated by any external agent, and due only to activity. It remains to use the fact that the denominator of the previous expression is not zero as a result of the definite sign of the dissipation. 
  Substitution of \eqref{velocity_gradient-eqn-reduced} into equation  {\eqref{n-eqn-reduce}} gives an ordinary differential equations on $\bn$.
  
Before proceeding, we summarize some properties of the function $g(2\varphi)$ that, in particular, highlight the role of the rotational  Leslie coefficient $\alpha_1$. 

\begin{proposition} 
Suppose that Parodi's relation \eqref{Parodi} holds. Then, the following statements hold:
\begin{enumerate}
    \item $g(2\varphi)=\text{constant}$ if and only if $\alpha_1=0$. 
    \item $g(2\varphi)$ is invariant under the transformation $\varphi\longrightarrow \pi-\varphi$ if and only if $\alpha_1=0$.
\end{enumerate}
\end{proposition}

\begin{remark}
 The vorticity $\partderiv{w}{x}$ has the same symmetry property as $g(2\varphi)$. 
\end{remark}
  \subsection { Director field pattern}
 Equation \eqref{ang-momentum} of balance of angular momentum reduces to the ordinary differential equation 
  \begin{align}
      0=2\Er^{-1}\varphi''+\frac{1}{2}\partderiv{w}{x}(\gamma_1-\gamma_2\cos{2\varphi}). \label{angle-nonlinear}
  \end{align}
  together with \eqref{velocity_gradient-eqn-reduced}, that upon its direct substitution gives the single  ordinary differential equation, 
  \begin{equation}
      0=2\Er^{-1}\varphi''+\frac{\mathcal A}{2}(\gamma_1-\gamma_2\cos{2\varphi})\sin(2\varphi)g^{-1}(\phi). \label{phiODE-0}
  \end{equation}
  Using inequality \eqref{g-bounded-below}, we can further write
\begin{equation}
      0=\varphi''+\frac{\mathcal A\Er}{4}\gamma_2(\frac{\gamma_1}{\gamma_2}-\cos{2\varphi})\sin(2\varphi)g^{-1}(2\varphi). \label{phiODE}
  \end{equation}
Equivalently,
   \begin{equation}\varphi''-\frac{\gamma_2}{4}\A\,\Er(-\frac{\gamma_1}{\gamma_2}\sin(2\varphi)+\frac{1}{2}\sin(4\varphi))g^{-1}(2\varphi)=0,  \label{angle-pattern0}  \end{equation}
   which, due to the negativity of $\gamma_2$, can be seen as a pendulum equation with two forcing terms.
  % \begin{equation}2\Er^{-1}\varphi''-\frac{\mathcal A}{2}(-\gamma_1\sin(2\varphi)+\frac{\gamma_2}{2}\sin(4\varphi))g^{-1}(\varphi)=0  \label{angle-pattern0\end{equation}
The equilibrium solutions of equation \eqref{phiODE} are 
\begin{align}
   & \varphi=\pm\frac{\pi}{2}, \quad \varphi=0, \quad \text{and the solutions of the algebraic equation } \label{0-eq}\\
    &\gamma_1-\gamma_2\cos{2\varphi} \label{nonzero-eq}
    =0.% \label{equilibria}
\end{align}
That is, 
\begin{equation} \varphi_0=\cos^{-1}[\frac{1}{\sqrt{2}}(1+\frac{\gamma_1}{\gamma_2})^{\frac{1}{2}}], \quad \varphi_1=\pi-\varphi_0. \label{equilibrium-nonzero} \end{equation}
For the  viscosity coefficients as in \eqref{viscosity-values}, we have $\varphi_0=\frac{\pi}{3}=1.0472 (\text{rad}), \,  \varphi_1= \frac{2\pi}{3}=2.0944 (\text{rad})$.
These correspond to angular values  $\varphi_\perp=\pm 0.5236 $ (rad). 
 The value of $\varphi_0$ estimated from the experiments is $\varphi_0= 1.0758 $ (rad),  with 
 ${\varphi_{0}}_{\perp}= 0.4950 $ (rad).
 We then expect that a full chevron profile will be approximated by the angles $\pm\frac{\pi}{3}$ or the periodic equivalents. 
This allows us to estimate the ratio,
\begin{eqnarray}
    \frac{\gamma_1}{ \gamma_2}=\frac{\alpha_3-\alpha_2}{\alpha_2+\alpha_3}=-0.5487.
\end{eqnarray}
It is easy to check that the equilibrium points \eqref{0-eq} are neutrally stable whereas $\pm \varphi_0$ and $\pm\varphi_1$ are saddle points. This statement is  justified  by the straightforward  calculation of the eigenvalues of the linearized equations about such points.  
\subsubsection{Linear oscillatory pattern} \label{linear pattern}
We now appeal to the linear analysis to approximate the frequency of the pattern. 
The pattern of the system is approximated by linearizing  the equation about $\varphi=0$ which predicts oscillatory solutions about the horizontal.  The linear equation is given by
\begin{equation}
    0=\varphi''+\frac{\A}{2}\Er(\gamma_1-\gamma_2)g^{-1}(0)\varphi.\label{frequency-small}
\end{equation}
Note that the properties $\gamma_1>0$ and $\gamma_2<0$ guarantees oscillatory behavior. 
   The (square) frequency of solutions is given by
\begin{equation} \label{linear-frequency}
    \omega_0^2=\frac{\A\Er}{2} \left(\gamma_1- \gamma_2\right) g^{-1}(0)
    =\frac{aL^2}{2K}\left(\gamma_1- \gamma_2\right) g^{-1}(0), %= \frac{a}{2}L^2 \Gamma.  \quad \Gamma:= \frac{\left(\gamma_1- \gamma_2\right) g^{-1}(0)}{2K},
\end{equation}
where in the last step, we applied the definitions \eqref{scaling3} of the dimensionless activity parameter $\A$ and the Ericksen number $\Er$. 
The expressions for the  dimensional  frequency becomes
\begin{equation}
  \omega^2:= \frac{\omega_0^2}{L^2}=\frac{a}{2K}\left(\gamma_1- \gamma_2\right) g^{-1}(0) %= {a}\Gamma= c\,\mathcal C_0\Gamma .
    \label{omega-ATP}
\end{equation}
The previous relation gives the square of the  spatial frequency as a linear function of the stress activity parameter $a$ \eqref{activity-concentration} depending on the ATP concentration,  in full agreement with the experimental results (Fig 3 (B), \cite{guillamat2016} shows  $\omega^2$ in terms of $\log c$). 
The profile of $\varphi$ is obtained by integration of \eqref{frequency-small} giving the general solution
\begin{equation}
    \varphi(x)= A \cos\omega_0x +B\sin\omega_0x,\label{linear-angle-profile}
\end{equation}
where $A$ and $B$ are arbitrary constants. The linearized profile of $w'(x)$  is then given by
\begin{equation}
w'(x) =2\A g^{-1}(0) \varphi= 2\A g^{-1}(0)(A\cos\omega_0x +B\sin\omega_0x)
    \end{equation}
From the graph (Fig 2(b), \cite{guillamat2016}), we observe that the maximum velocity $w$ occurs at $x=0$, so 
$w'(0)=0$ (also confirmed by the vorticity graph of the same figure), justifying the choice  $A=0$ in the previous relations. Hence
\begin{equation}
     \varphi(x)= B\sin\omega_0x, \quad w'(x)=  2\A g^{-1}(0)B\sin\omega_0x, \label{linear-phi-profile}
\end{equation}
and 
\begin{equation}
    w(x)=-\frac{2\A}{\omega_0} B\cos\omega_0x.
\end{equation}
Denoting $V_{\text{max}}>0$ the  maximum (dimensional) speed, from Figure Figure 3C) \cite{guillamat2016}, we infer the linear relation $V_{\text{max}}
=1.7\alpha$ as well as the concentration and velocity data summarized in table 2. We now choose the typical velocity $V$ appearing in the scaling relations 
\eqref{scaling1}-\eqref{scaling3} to be $V=V_{\text{max}}= 4\mu {\text m/s}, $ a mid-range value. This gives 
\begin{equation} B=-\frac{\omega_0}{2\A}=-\frac{\omega L}{2\A } \quad \text{or}\quad B=   -\frac{\omega_0 V_{\text{max}}}{2aL}\eta=-\frac{\omega V_{\text{max}}}{2a}\eta, \label{B-expression}
\end{equation}
where we have used \eqref{scaling3} for $\A$.  An alternate representation gives
\begin{equation}
    B=-\frac{1}{2}\sqrt{\frac{\Er}{2\A} (\gamma_1-\gamma_2)g^{-1}(0)}.
\end{equation}
\begin{table}[h!]
\centering
%\begin{center}
\begin{tabular}{ |c|c|c|c|} 
 \hline
 $\alpha=\ln c$&$V_{\text{max}} (\mu$m/s)&$ {\lambda_0}^{-2}(\mu\text{m})^{-2}$ \\
 \hline
4.6 & 2 & -\\ 
5.5& -& $10^{-4}$\\
 5.8 & 4  & - \\ 
 6.3& -& $2*10^{-4}$\\
 7& 6 & -\\ 
 7.1&- & $3*10^{-4}$\\
 \hline
\end{tabular}
\caption{These are values obtained from  Figure 3C) \cite{guillamat2016}.
The units for the velocity are $\mu$m/s and $c$ is measured in [ATP]/$\mu$m, where $[\cdot]$ denotes concentration.  $\lambda_0$ denotes the line spacing and corresponds to a half wave length. $L=5*10^3\mu$m, $\eta=\frac{4}{3}*10^{-2} \frac{\text g }{\mu{\text m}*{\text s}^2}$.  }
\end{table}
 To fully determine $B$, it remains to estimate $\omega$ and the stress parameter $a$. For this, 
 we first obtain $\omega$ from the graph  in  Figure 3 (B), \cite{guillamat2016}, which gives the relations 
 \begin{equation}
 \omega^2=\big(\frac{\pi}{\lambda_0})^2=  s_0\,\alpha\, ({\mu{\text m}}^{-2}),  \quad \lambda_0=\frac{\pi}{\sqrt{s_0}}\alpha^{-\frac{1}{2}}\, \mu{\text {m}}\quad  \text{with}\quad s_0=0.8\pi^2 10^{-4} \label{omeaga-line}
 \end{equation}
where we have used the value $0.8*10^{-4}$ for the measured slope of the line $\lambda_0^{-2}$ versus $\alpha$ (Table 2).  These yield the  bounds 
\begin{equation}
 {\mu\text{m}}^{-1}   3.9738*10^{-2}\leq \omega\leq 7.4343*10^{-2} {\mu\text{m}}^{-1}, \quad   79.0569\,{\mu\text{m}}\,\geq \lambda_0\geq 42.2577\,{\mu\text{m}}\label{omega-bounds}
\end{equation}
for the $\alpha$-range  $2\leq\alpha\leq 7$. 
%\begin{remark}   The experimental measurements of the line separation give the average value  $\lambda_0\approx 35.25$ $\,\mu m$, which corresponds to   $\omega=\frac{2\pi}{\lambda}=0.04472 \,$ $ {\mu \text{m}}^{-1}.$ These values fall within the bounds \eqref{omega-bounds}, for $\alpha=2.53.$\end{remark} 
In a related calculation, we combine relations \eqref{omega-ATP} with \eqref{omeaga-line} giving 
\begin{equation}
    a= \frac{2K s_0 g(0)\alpha}{\gamma_1-\gamma_2} \label{a-alpha}
    % = \frac{\omega^2{\ba4\Gamma}.
\end{equation}
A subsequent  comparison with \eqref{activity-concentration} yields
\begin{equation}
C_0=\frac{2Ks_0g(0)}{\gamma_1-\gamma_2} \alpha e^{-\alpha}.
\end{equation}
The latter predicts that the most efficient rate of conversion $C_0$ occurs when the chemical activity is $\alpha=1$.
{So, for $4\leq \alpha\leq 7$, the experimental regime of the chemical activity parameter, the factor $C_0$ falls into the decaying regime. This could be an artifact of the linear approximation of the flow. However, from equation \eqref{a-alpha}, we see that the stress activity parameter is linear in $\alpha$. }
\begin{table}[h!]
\centering
%\begin{center}
\begin{tabular}{ |c|c|c|c|c|c|} 
 \hline\hline
 $\text{Quantity}$& wave number & lane spacing &chem activity& chevron angle& Activity \\
 \hline
Symbol& $\omega$ & $\lambda_0$ & $\alpha$ & $\varphi_0=|B|$  & $\A$ \\ 
 Value & 0.0447&  $70.25$ & [2, 7] & $1.0758$ & $0.1735*10^3$ \\ 
 Units& ${\mu m}^{-1}$ & $\mu$m & 1 & rad & 1\\ 
 \hline
\end{tabular} \label{table line-spacing}
\caption{The values shown here correspond to the experimental measurements. The dimensionless activity $\A$ results from such values.  The range of $\alpha=\log c$ is also the experimental one for the chemical activity. }
\end{table}
We now proceed to calculating the viscosity ratio
$\frac{g(0)}{\gamma_1-\gamma_2} $ by application of the second relation in \eqref{B-expression}:
Substituting the values previously obtained into that expression, we get
\begin{equation}
    |B|=1.0758= \frac{\omega V_{\begin{tiny} max\end{tiny}}}{2a}\eta= \frac{0.00224*4}
    {2*\frac{2Ks_0g(0)\alpha}{\gamma_1-\gamma_2}} \eta.
\end{equation}
This relation, together with \eqref{alpha2-negative} gives
\begin{eqnarray}
    \alpha\frac{g(0)}{\gamma_1-\gamma_2} =0.3516 \quad \text{and}\quad \alpha_2=-\frac{\alpha}{0.7032} g(0). \label{al-B}
\end{eqnarray}
Likewise,   the relation $\gamma_2=\alpha_2+\alpha_3$ gives
\begin{equation}
    \gamma_2=-1.8365\alpha. \label{bar-gamma2}
\end{equation}
A straightforward calculation using \eqref{g-function} gives
\begin{equation}
g(0)=\alpha_2+\alpha_4+\alpha_5. 
\end{equation} 
Taking the combination $g(0) $ as the effective viscosity of the system, we set
\begin{equation}
    g(0)= \alpha_2+ \alpha_4+\alpha_5=1, \label{effective-viscosity}
\end{equation}
This relation combined with \eqref{alpha2-negative} and \eqref{al-B} gives
\begin{equation}
    \gamma_1-\gamma_2= \frac{\alpha}{0.3516}\quad \text{and}\quad \alpha_2=-\frac{\alpha}{0.7032}.
\end{equation}
Taking into account the experimental range of values  of the activity  $4\leq \alpha\leq 7$, we find that
\begin{equation}
    2.8441\leq |\alpha_2|\leq 9.9545.
\end{equation}
Using the relation
\begin{equation}
    \cos{2\varphi_0} =-0.5487=\frac{\alpha_3-\alpha_2}{\alpha_3+\alpha_2}.
\end{equation}
yields
\begin{equation}
    \alpha_3=0.2914\alpha_2=-0.4144\alpha. 
\end{equation}
The previous relation together with \eqref{effective-viscosity} yields
\begin{equation}
    \alpha_4+\alpha_5= 1-\alpha_2=1+ 1.4221\alpha.
\end{equation}
Likewise, following equation \eqref{a-alpha}, we get 
\begin{equation}
    a=5.55*10^{-5} (\text {stress units}).
\end{equation}
Next, we estimate the  dimensionless parameters  $\A$ of the flow. Using the definitions of $\Er$ and $\A$ given by  together by \eqref{scaling3} together with \eqref{omeaga-line} and the previously calculated expressions of $\gamma_1$ and $\gamma_2$, that is, 
\begin{equation}
    \A=\frac{a L}{\eta V}= \frac{2 K s_0 g(0)}{(\gamma_1-\gamma_2)\eta V }L\alpha =\frac{2  s_0 g(0)\alpha}{(\gamma_1-\gamma_2) }\frac{L^2}{\Er}=
    \frac{2   g(0)\omega^2}{(\gamma_1-\gamma_2) }\frac{L^2}{\Er}
    %\frac{\omega^2 L}{\eta V\Gamma}= \frac{\log c \lambda}{\eta V\Gamma}.
\end{equation}
This gives 
\begin{equation}
    \A= \frac{2*1.4*10^{-4}}{-2\alpha_2*80}L^2\alpha=0.1735* 10^3.
\end{equation}
Recalling the earlier calculation giving $\Er=80$, we arrive at 
\begin{equation}   \A\Er = 1.388*10^4. \label{A*E}\end{equation}
 The fact that $\Er <\A$ indicates that the flow is dominated by the activity over the viscous forces,  in contrast to the the case involving polymers. However, we observe from equation \eqref{frequency-small} that it is the product $\A\Er$ what drives the pattern formation. 
\begin{table}[h!]
\centering
%\begin{center}
\begin{tabular}{ |c|c|c|c|c|c|c|} 
 \hline\hline
 $\text{Leslie coefficient }$& $\alpha_2$ & $\alpha_3$ & $\alpha_4+\alpha_5$& $\alpha_6$& $\frac{\gamma_1}{\gamma_2}$& $\gamma_2=\alpha_2+\alpha_3$ \\
 \hline
 Value & $-1.4221\alpha$ &  $-0.4144\alpha$ &$1+1.4221\alpha$  & $\alpha_5-1.8365\alpha$ &-0.5487& $-1.8365\alpha  $\\ 
 \hline
\end{tabular} 
\caption{The values of the viscosity coefficients as functions of $\alpha$, $4\leq \alpha\leq 7.$ Here, we assume $\alpha_1=0$. For specific calculations, we take the average value $\alpha=5.5$ }\label{viscosity-table}
\end{table}
%\begin{remark} Apart from the representative value of $\omega$ given by \eqref{omega-estimate},  we can obtain its experimental range by further exploration of the graphFigure 3(B)  \cite{guillamat2016} that gives
%\begin{align*}0.2*10^{-4}{\mu \text{m}}^{-2}<\lambda_0^{-2}<4*10^{-4} {\mu \text{m}}^{-2}\\1.97*10^{-4}{\mu \text{m}}^{-2}<\omega^2=\frac{\pi^2}{\lambda_0^2} < 39.5*10^{-4}{\mu \text{m}}^{-2}.\end{align*}
%The latter inequalities combined with \eqref{A*E} give
%\begin{align}  0.8*10^4<\A\Er<1.6*10^5, \quad  10^2<\A<2*10^3.\end{align}\end{remark}
%\begin{remark} Recalling that the amplitude of $\varphi$ should be within the range $\frac{\pi}{3}$ to $\frac{\pi}{6}$ as indicated by the equilibrium value of $\varphi$, we see that the $0.1118$ radians obtained here fall short of the former. This could be due to the linear analysis. 

%\end{remark}
\subsubsection{Nonlinear oscillatory pattern} 
The governing equation \eqref{angle-pattern0} in the case $\alpha_1=0$ becomes
 \begin{equation}\varphi''-\frac{\gamma_2}{4}\A\,\Er(-\frac{\gamma_1}{\gamma_2}\sin(2\varphi)+\frac{1}{2}\sin(4\varphi))=0.  
 \label{angle-pattern0-simple}  
 \end{equation}
 We observe that, in view of the data in table \ref{viscosity-table} and equation \eqref{A*E}, the solutions of this equation depend on the single activity parameter $\alpha$.  Moreover, the equation
 \eqref{A*E} together with  \eqref{bar-gamma2} give
 \begin{equation}\label{def_espilon}
   \epsilon^{-2}:=  \frac{|\gamma_2|}{4} \A\Er= 0.6373*10^4\alpha, \quad 2\leq \alpha\leq 7.
 \end{equation}
 An immediate consequence of equation \eqref{angle-pattern0-simple} is the scaling property that confirms the experimental results in Fig 2B \cite{guillamat2016}. 
 The change of variables in \eqref{regular+internal}  $ x\longrightarrow \frac{x}{\epsilon} $ transforms equation \eqref{angle-pattern0-simple} into 
  \begin{equation}\frac {d^2\varphi}{d x^2}+(-\frac{\gamma_1}{\gamma_2}\sin(2\varphi)+\frac{1}{2}\sin(4\varphi))=0.  
  \label{angle-pattern-base}  \end{equation}
  \begin{figure}
  %\centering
\begin{center}
 \begin{subfigure}[b]{.47\textwidth}
\quad \quad \quad \quad \quad \quad \includegraphics[height=3.9cm]{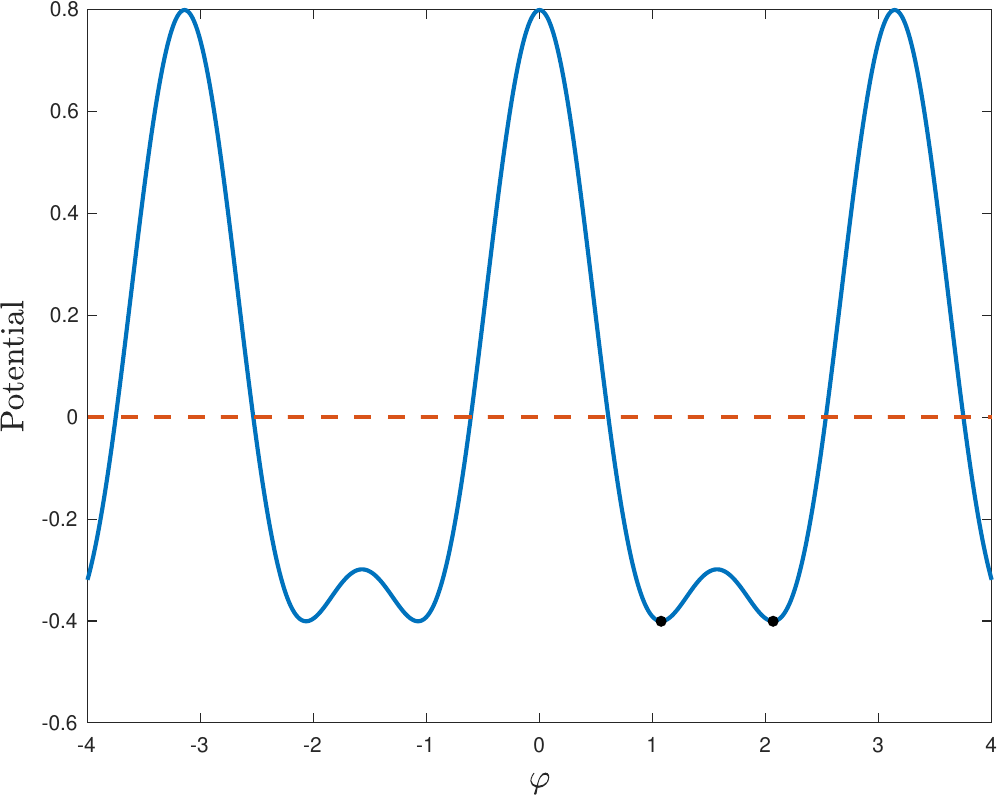} 
\caption{Potential}
\label{fig:potential}
 \end{subfigure}
\hfill
 \begin{subfigure}[b]{.47\textwidth}
\quad \quad\includegraphics[height=4.2cm]{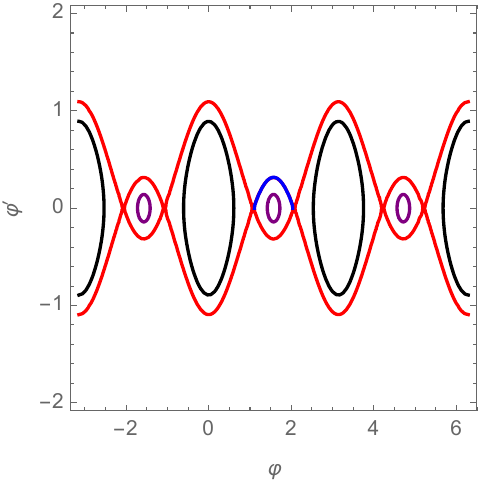} 
\caption{Trajectory in Phase plane}
\label{fig:phaseportrait}
 \end{subfigure}
 \caption{(a) Graph of the potential energy \eqref{Potential}. The critical points  $\varphi_0$ and $\varphi_1$ correspond to consecutive chevron angles. These values are  saddle points of equation \eqref{angle-pattern0}. \,(b) Trajectory in the phase plane $(\varphi,\varphi')$ from \eqref{angle-pattern-base}.}
 \label{Potential-fig}
 \end{center}
%\caption{Plots of the potential (a),  a sketch of the phase plane of the nonlinear equation (b). }
\end{figure}
The Sturm-Liouville theory determines that the previous equation has periodic solutions \cite{hale2009ordinary}.
The distance between two consecutive maxima corresponds to the scaled dimensionless line separation, $\lambda_{00}$, of equation \eqref{angle-pattern-base}. Note that the line separation $\lambda_{0}=O(1)$ is independent of $\alpha$. 
Changing back to the original (dimensional) variable $x$, it gives 
\begin{equation}
    \lambda_0= \epsilon \lambda_{0}L= \frac{2}{\sqrt{|\gamma_2|\A\Er} }\lambda_{0} L
 =1.4785\frac{\lambda_{0}L}{\sqrt{\A\Er\alpha}}=1.9697*10^2\frac{\lambda_{0}}{\sqrt{\alpha}}, \label{line_separation-nonlinear}
\end{equation}
  where we have used tables \ref{experimental-values} and \ref{viscosity-table}, and equation \eqref{A*E}. We note that this scaling agrees with that derived from experiments as shown in figure 2B \cite{guillamat2016}.  Numerical simulations of the governing equation yield the graphs in figure 6. 

Now, we turn into further exploring the properties of equation \eqref{angle-pattern0-simple}.
 
Multiplication of  equation \eqref{angle-pattern0-simple} by $\varphi'$:
\begin{equation}
\epsilon^2 \varphi' \varphi''+\varphi'(-\frac{\gamma_1}{\gamma_2}\sin(2\varphi)+\frac{1}{2}\sin(4\varphi))=0,
\end{equation}
    and integrating with respect to $x$ we obtain the 1-parameter family of first integrals
    \begin{equation}
       \epsilon^2  \varphi'^2-\big(\frac{\gamma_1}{|\gamma_2|}\cos{2\varphi}+\frac{1}{4}\cos{4\varphi}\big)=\text{Const} \label{first-integral}
    \end{equation}
Applying the values from table 4, we get
\begin{equation}
\epsilon^2\varphi'^2-(0.5487\cos{2\varphi} +
\frac{1}{4}\cos{4\varphi})= \text {Const}, \quad 4\leq \alpha\leq 7. \label{first-integral-special}
\end{equation}
Note that \eqref{first-integral} and \eqref{first-integral-special} involve the potential energy
\begin{equation}
    \text{Potential}:= \frac{\gamma_1}{|\gamma_2|}\cos{2\varphi}+\frac{1}{4}\cos{4\varphi} \label{Potential}
\end{equation}
As stated in section \ref{linear pattern}, the equilibrium solutions of the problem are given in equations \eqref{0-eq} and \eqref{nonzero-eq}, the first ones corresponding to centers of equation \eqref{angle-pattern0-simple} and the latter ones saddle points. 
Figures 3 (a) show the the potential energy \eqref{Potential} and  a plot of  phase orbits. Note that, they both depend on the ratio $\frac{\gamma_1}{\gamma_2}$  only. A whole periodic orbit around the center $\varphi=\frac{\pi}{2}$ in the phase-plane is shown in figure 3 (b).

Figure 6 gives the calculated separation distance values of the chevron with respect to the activity parameter $\alpha$. 

A numerical simulation based on the heteroclinic orbit gives the (dimensionless) distance $\bar\delta=6.980*10^{-5}$ separating the ends of the intervals with $\varphi=\varphi_0$ and $\varphi=\varphi_1$. (This is calculated for $\alpha=5.5$). In terms of the dimensional variables, \begin{equation}\delta=6.988*10^{-5}*L= 0.35 \mu\text{m}.  \label{delta}\end{equation} This line separation $\lambda_0$ compared to the former half chevron length $\lambda_0$  gives  $\frac{\lambda_0}{\delta}= 2*10^2.$ A plot of $\frac{1}{4}$ of the heteroclinic orbit use in the calculation  of $\delta$ is shown in figure 12 of the Appendix. 

Plots of the  solutions of the nonlinear equations \eqref{velocity_gradient-eqn-reduced} and \eqref{angle-pattern-base} for the angle of director alignment  and  the velocity field are  shown in figure 7.
%in the interval $[4, 7]$.
%In terms of the numerical values of in the tables,  the governing equation becomes
%\begin{equation}  \varphi''+ 3.91*10^3\alpha(0.5487\sin{2\varphi}+ \frac{1}{2}\sin{4\varphi})=0.\end{equation}
\begin{figure} \label{lane_spacing-activity}
\begin{center}
% \begin{subfigure}[b]{.60\textwidth}
  \includegraphics[width=0.35\textwidth]{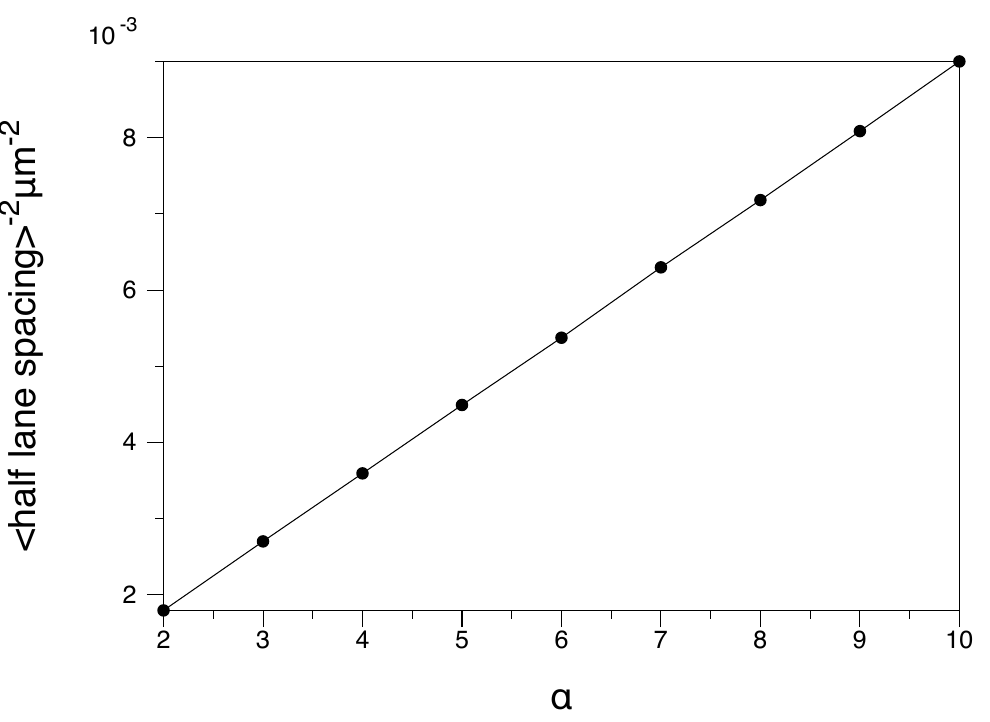}\quad \quad \quad \quad 
  \includegraphics[width=0.25\textwidth]{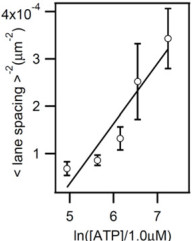}
  \caption{This line represents the inverse square of the half line versus the chemical activity $\alpha$, resulting from the numerical solutions of the nonlinear problem. The parameters are based on table \ref{viscosity-table} with $\alpha_4=4/3\times 10^{-3}$ and the initial value $\varphi_0=1.0758$. }
 %\end{subfigure}
 \end{center}
\end{figure}

\begin{figure}
\begin{center}
 \begin{subfigure}[t]{.495\textwidth}
 \begin{center}
  \includegraphics[width=1\textwidth]{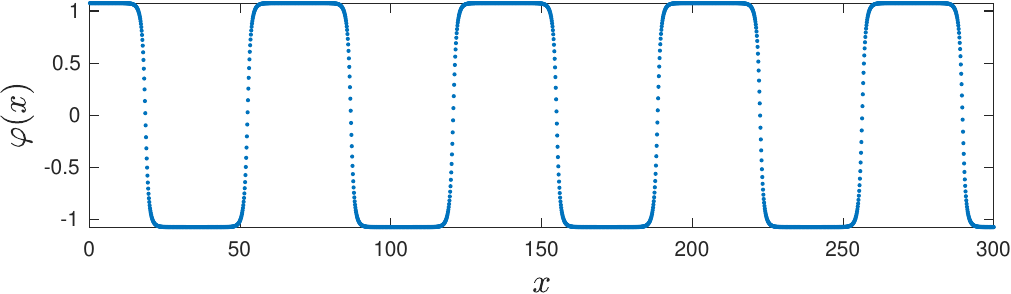} 
  \caption{Profile of the angle of director alignment}
  \bigskip
   \end{center}
 \end{subfigure}
 \begin{subfigure}[t]{.495\textwidth}
 \begin{center}
  \includegraphics[width=1\textwidth]{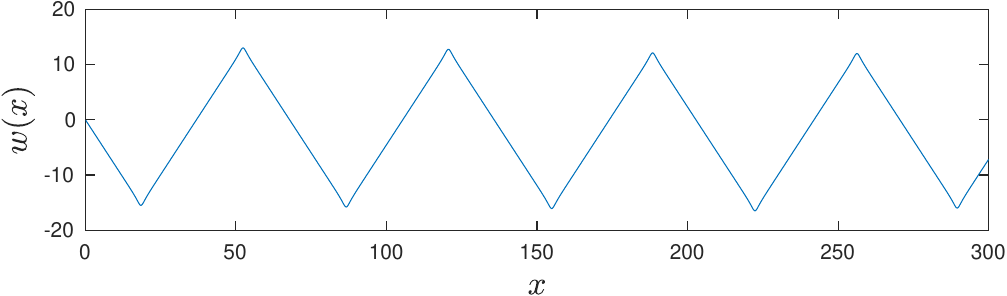} 
  \caption{Profile of the velocity field}
   \end{center}
 \end{subfigure}
 \caption{Numerical simulations of the  velocity field (b)  and directional angle (a) obtained as solutions of  equations  \eqref{velocity_gradient-eqn-reduced} and \eqref{angle-pattern-base}, respectively. The spatial variable $x$ is in $\mu$m units, $w(x)$ in $\frac{\mu m}{\text{sec}}$ and $\varphi$ in radians.  Parameters are selected as $\alpha=2$,     $\alpha_1=0$, 
$\alpha_2=-2.8442$, $\alpha_3=-0.8288$, $\alpha_4=0.0013333$, $\alpha_5=3.8428667$,  $\alpha_6=-3.1453$, and $\gamma_1/\gamma_2= -0.5487$ in  Table \ref{viscosity-table}.} 
\end{center} 
\end{figure}
 
 \subsection{Internal layer structure} \label{internal-layer}

%In  section \eqref{linear pattern}, we identified parameters of the problem by matching the experimental results on the wave length and angular parameter of the pattern with the predictions from the linear theory, that is, accounting for an oscillatory pattern about the equilibrium $\varphi=0$.

In this section, we use asymptotic arguments to probe the pattern in terms of the boundary layer theory. Taking into account  that   the product $\A\Er>>1$ \eqref{A*E} (in fact, it is of the order $10^4$),  the governing equation \eqref{angle-pattern0-simple} is singularly perturbed so the solutions are expected to show  internal layer transitions. That is, the profile of $\varphi(x)$ will be alternately close to the equilibrium values  $\varphi_0$ and $\varphi_1$, respectively, given by \eqref{equilibrium-nonzero} in most of the flow domain, and exhibiting  sharp transitions between such values.
This is consistent with the banding pattern observed in the flow, such as that shown in Fig 1 (D, H, J) \cite{guillamat2016}.
%(In figure J, the direction of the $x$-axis corresponds to that of the white bar). 

Let us now outline the asymptotic structure of the solution. 
Let the small parameter $\epsilon>0$  be as in  \eqref {A*E}.
We consider a family of $n$ intervals of length $\lambda_0$, the line spacing defined in section  \eqref{linear pattern}, and the collection of end-interval points, $\{k\lambda_0\}_{k=-n, -n+1, \dots, 0, 1, \dots n-1, n}$, with $n=[\frac{L}{2\lambda_0}]$, where the brackets denote the largest positive integer smaller than $\frac{L}{2\lambda_0}$, the $L$ representing the horizontal length of the domain. 
Note that, in the case that the bracketed quantity is a positive integer, that is $L=2n\lambda_0$,  the domain contains $n$ full chevron units of length $2\lambda_0$ each. We consider a piecewise constant function 
$$ \Phi(x)=  \begin{cases} \varphi_0, \, &x\in (-n \lambda_0, (-n+1)\lambda_0)\cup ((-n+2) \lambda_0, (-n+3)\lambda_0)\dots \cup  ((n-2)\lambda_0, (n-1)\lambda_0)\\
\varphi_1, \,& x\in ((-n+1)\lambda_0, (-n+2)\lambda_0)   \cup ((-n+3)\lambda_0, (-n+4)\lambda_0)  \dots\cup((n-1)\lambda_0, n\lambda_0)
\end{cases}$$
that is, taking  values $\varphi_0, \varphi_1$, respectively, on alternating intervals of length $\lambda_0$. For convenience, we take $n$ to be odd, so that the one-wave length section of the profile
\begin{equation}\Phi(x, \epsilon)= \begin{cases} \varphi_0, \, &x\in(-\lambda_0, 0)\\
\varphi_1, \, &x\in (0, \lambda_0), \end{cases} \quad \quad  [\varphi]:= \varphi_1-\varphi_0.\label{center-pattern}
\end{equation}
 is located at the center of the pattern. It then repeats to the right and to the left, along an interval of length $L$, with the neighboring  cells begin connected by  smooth functions matching the constant values $\varphi_0$ and $\varphi_1$. 
 With a small modification on the notation, we represent a solution  of   equation \eqref{angle-pattern0-simple} as 
\begin{equation}
    \varphi(x,  \tilde x, \epsilon)= \Phi(x, \epsilon) + H(x) I\Phi( \tilde x, \epsilon), \quad  \tilde x=\frac{x}{\epsilon}\geq 0, \label{regular+internal}
\end{equation}
where $H$ denotes the Heaviside function. Although $x\in (-\lambda_0, \lambda_0)$,  the stretch variable $ x$ is defined only for $x\geq 0$. Following standard procedure, we assume that the terms $\Phi(x, \epsilon)$ and  $I\Phi( \tilde x, \epsilon) $ admit  regular asymptotic expansion with respect to the small parameter $\epsilon$.

%(Without loss of generality, we focus on the wave-length section in the central interval $(-\lambda_0, \lambda_0)$).
The leading term of the regular asymptotic expansion $\Phi(x, 0)$ solves the governing equation with $\epsilon=0$, that is, it subsequently takes the values $\varphi_0, \varphi_1$ as in \eqref{center-pattern}.  To obtain the equation for the 
leading term $I\Phi(\bar x, 0)$, we define 
$$G(2\varphi):= (\frac{\bar\gamma_1}{\bar\gamma_2} -\cos2\varphi)\sin2\varphi g^{-1}(2\varphi),$$
and perform the  Taylor approximation about $\Phi(x, 0)$ about $\varphi=\varphi_1$,
 \begin{equation} G(2\varphi)= 2\sin^2(2\varphi_1)g^{-1}(2\varphi_1)(\varphi-\bar\varphi_1)+O((\varphi-\varphi_1)^2).\end{equation}  
Denoting 
 \begin{align}
       \omega^2_1:= & 2\sin^2(2\varphi_1) g^{-1}({2\varphi_1})= 2(1-\frac{\bar\gamma_1^2}{\bar\gamma_2^2})  g^{-1}(2\varphi_1), \label{frequency-large}
   \end{align}
the equation for the leading internal layer term is 
\begin{equation}  0=\frac{d^2(I\Phi)}{d\tilde x^2} - \omega_1^2I\Phi, \quad I\Phi(0)=\varphi_0-\varphi_1, \quad \lim_{\epsilon\to 0}I\Phi(\frac{x}{\epsilon}) = 0. \end{equation} 
    The solution to the previous boundary value problem is
    \begin{equation}
        I\Phi(x, 0)= (\varphi_0-\varphi_1) e^{-\omega_1\frac{x}{\epsilon}}. \label{transition}
    \end{equation}
   Hence, the central pattern solution  takes the form
   \begin{equation} \label{A-chevron}
       \varphi(x, \frac{x}{\epsilon}, \epsilon)= \begin{cases}  \varphi_0+O(\epsilon),  & -\lambda_0\leq x\leq 0, \\
        \varphi_1+ (\varphi_0-\varphi_1)  e^{-\omega_1\frac{x}{\epsilon}},  & \lambda_0\geq x\geq 0. \end{cases}
   \end{equation}
\begin{remark}
Note that \eqref{A-chevron} represents the solution of a unit chevron transitioning from the values $\varphi_0$ to $\varphi_1$, from let to right of the interface. The representation corresponding to the transition from $\varphi_1$ to $\varphi_0 $ has the same form but interchanging  $\varphi_0$ and $\varphi_1 $
 in the expression. 
\end{remark}
We now give an estimate of the thickness $\delta_{BL}$ of the transition region between $\varphi_0$ and $\varphi_1$, appealing to the {\it half-life} concept of exponential decay.  So the thickness corresponding to the angular decay from $[\varphi] $ to $\frac{1}{2}[\varphi]$  is easily calculated as 
\begin{equation}
    \delta_{BL}= \epsilon\frac{\log{2}}{\omega_1} \mu\text{m}.
\end{equation}
Furthermore, using the values in the values of $\varphi_1$ and $\gamma_1$ from the tables, we  calculate $\omega_1^2$ as 
\begin{equation}
    \omega_1^2=|\bar\gamma_2|\sin^{2}{2\varphi_1}= 0.0137\alpha,
\end{equation}
and 
\begin{equation}
    \delta_{BL}= 1.585*10^{-2}\alpha^{-\frac{1}{2}}  \mu\text{m}.\label{delta-asymptotic}
\end{equation}
We point out that the size of the transition region decreases with increasing chemical activity, resulting in a sharper structure. 
Taking the mid-range value $\alpha=5.5$, we find that $\delta_{BL}=0.660*10^{-2}\mu \text{m}. $
Note that this value is significantly smaller than that predicted from the value $\delta= 0.35 \mu\text{m}$ in \eqref{delta} as calculated from the heteroclinic orbit. This may correspond to the fact that, although the actual value $\varepsilon= O(10^{-4})$ is small, it does not quite conform to the 0-limit.

\section{Optimal rate of energy dissipation}
We now adopt a different point of view and investigate whether the amplitude and wave length of the pattern can be selected by minimizing the total rate of dissipation function \eqref{dissipation-total}. 
%\begin{eqnarray}\Delta_N = \frac{1}{2}\left(\alpha _1(n\cdot An)^2+\gamma _1 \left|N|^2+\left(\alpha _2 +\alpha_3+\gamma_2 )\right(N\cdot A n)+\alpha _4 |A |^2+(\alpha_5 +\alpha _6\right)\left| A n \right|^2\right) \label{dissipation-rate}\end{eqnarray}
 For the flow as in \eqref{nematic-reduced-form}, and using equations \eqref{velocity_gradient-eqn-reduced} and \eqref{g-function},  it  reduces to 
\begin{gather}
   \mathcal R= M_0 \int_0^1\Delta_N(\bar\alpha_i, \bar w_{\bar x}) \,d\bar x, \quad \text{with} \label{total-dissipation-scaled} \\ 
\Delta_N(\bar\alpha_i, \bar w_{\bar x}(\bar x))=\sin^2{2\varphi} \frac{R(2\varphi)}{g^2(2\varphi)}, \\
      R(2\varphi):=  \bar\alpha_1\sin^2(2 \varphi)-2(\bar\alpha_2+\bar \alpha_3) \cos(2\varphi )
+2(\bar\alpha_3+\bar\alpha_4+\bar\alpha_5),
\end{gather}
 %(Explicit calculations of the terms in  \eqref{dissipation-rate} are given in the Appendix). 
 $M_0>0$ represents the total rate of dissipation scale constant $M_0= \text {Vol}(\Omega_N) \frac{V^2}{L^2}\eta .$
We note  that 
 $ R(2\varphi)= R(2(\pi-\varphi))$, but $g(2\varphi)\neq g(2(\pi-\varphi))$ (unless $\alpha_1=0). $ 
%\subsection{Evaluate rate of dissipation on linear oscillatory profiles} 
Denoting $B>0$ the average angle of alignment of the chevron, 
we approximate the total rate of dissipation $\mathcal R$ using the linear solution profile \, $\varphi(x)=B\cos{\omega x}$   \eqref{linear-phi-profile}.
Upon substitution into \eqref{total-dissipation-scaled}, it gives
$\mathcal R=\mathcal R(B, \omega),$  a nonconvex, oscillatory, function  of $(B, \omega) $
showing multiple potential wells.  We seek approximations to the critical points of $\mathcal R=\mathcal R(B, \omega)$  in two particular cases.  
\smallskip

\begin{figure}  
\begin{center}
\includegraphics[width=0.4\textwidth]{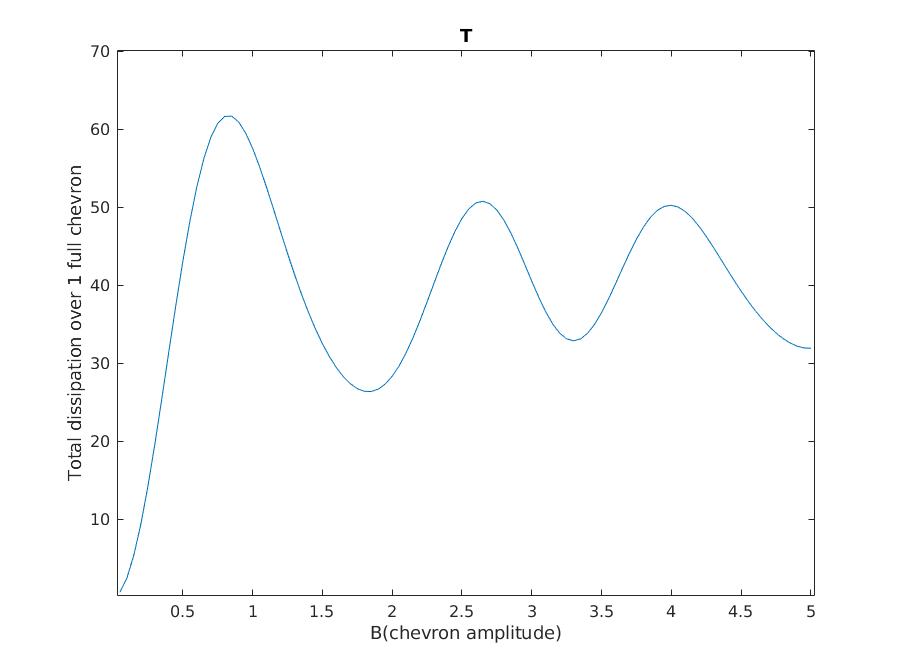}\quad 
\includegraphics[width=0.4\textwidth]{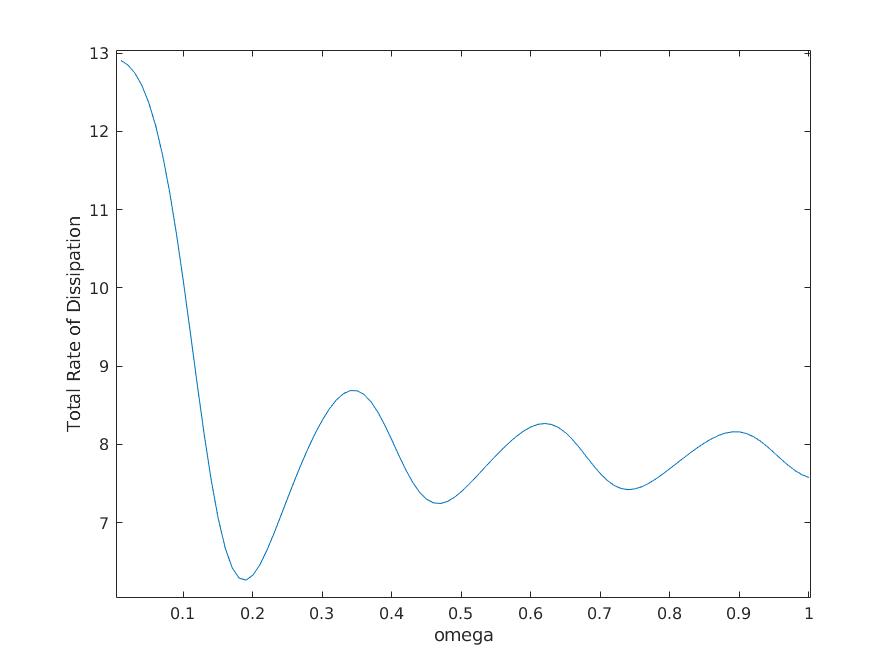}
\caption{Left: We set $\omega=0.177$ as in experiment. The first nonzero minimum for the angular amplitude of the chevron corresponds to $\varphi\approx 1.98$, and the half angle with respect to the vertical  $\varphi_\perp=0.4092$ (rad).  Right: We set the angular amplitude of the chevron to $B=1.052$ approximately as in one of the equilibrium.  The first nonzero minimum for the chevron wave length  is $\omega\approx 0.18$ ${\mu\text{m}}^{-1}. $}\label{Calc-TRD}
\end{center}
\end{figure}

{\bf 1.\,}  We fix $\omega$ within the experimental range and find the first nonzero minima of $\mathcal R(B)=\mathcal R(B, \omega_|\text{\tiny{fixed}}).$ 
The outcome of such a minimization for $\omega=0.177 \mu$m is shown in figure \ref{Calc-TRD} (left). The minimum value at $B=1.8$ satisfies $\frac{\pi}{3}<1.8<2*\frac{\pi}{3}$, placing it between the constant equilibrium  values \eqref{equilibrium-nonzero}. The chevron semi-angle with the vertical is found at 
$\varphi_\perp\approx 0.4950$ (rad).  Figure \ref{Calc-TRD} (lelft) illustrates this finding. 
\smallskip

{\bf 2.\,} Setting the angular amplitude $B= 1.052$ (rad), we find that the wave length that corresponds to the first nonzero minimum of the rate of dissipation function corresponds to $\omega\approx 0.18 $ ${\mu\text{m}}^{-1}. $ This outcome is illustrated in figure \ref{Calc-TRD} (right). 
%Fix $B$ as in equilibrium value. Minimize $\mathcal R$ with respect to $\omega$.
\begin{remark} In section \eqref {RateDissipationExp} of the Appendix,  we have  calculated the total rate of dissipation for the boundary layer asymptotic representations of the pattern given by \eqref{regular+internal}. We found that $\mathcal R$ is an increasing function of the wave length, and so, it is unable to select a preferred wave length.  This follows from the fact that  the exponentially thin transition regions between chevron domains  do not pose a sufficiently large penalty to uncover a wavelength.

\end{remark}

%%%%%%%%%%%%%%%%%%%%%%%%%%%%

\section{Conclusions}
We have studied the  flow of an active nematic liquid crystal confined within a smectic A environment. Guided by highly precise experimental results, we have applied several analytic approaches to characterize the wavelength, angle of alignment and flow velocity of the emerging chevron pattern.  We have first used  linear analysis as a data fitting mechanism, followed by the nonlinear approaches based on the phase plane and the asymptotic method of  boundary layers to determine how different aspects of the pattern depend on the activity parameter.    We concluded  the work exploring how   the application of the principle of minimum dissipation can be also applied to determine the amplitude and wave length of the pattern.  

This work highlights how the combination of analytic approaches with available data can  deliver a comprehensive characterization of material behavior.   It is well known that one of the difficulties in modeling liquid crystal flow is the lack of knowledge of the  Leslie  coefficients. We have illustrated how to calculated special combinations of such coefficients, including the Miesowicz viscosities.  Further exploration of the analytic methods may lead to a full individual characterization of such coefficients and their dependence on the activity parameter. 

One of the outcomes of the work  is the ability to compare analytic and experimental results to infer values of the Leslie coefficients form the active nematic. 

This work also brings some new light on the assumptions on the viscosity coefficients that lead to well-possedness of the Ericksen-Leslie  model, beyond Leslie's inequalities, a problem that has remained open for several decades. In future analysis we will explore the role of the Miesowicz viscosities on guaranteeing existence of solutions  the anisotropic hydrodynamic model. 

%In particular, we justify the choice $\alpha_1=0$,  usually a very difficult parameter to measure, because, otherwise, a symmetry breaking would be observed in the pattern. In particular, $\alpha_1\neq 0$ predicts different speeds between the Chevron channels traveling in opposite directions, property that has not been experimentally observed. 

%\noindent\(\pmb{\text{Plot}[ \text{NIntegrate}[\text{delta}[x,H,0],\{x,0, 4*\pi /\text{w0}\}],\{H,0,20\}]\text{(*ac=1*)}}\)

\bibliographystyle{plain}
\bibliography{active_matter_ref}

\newpage
\appendix 
 \section{}
 
 \subsection{The governing equations in dimensionless form} \label{ApndixA}
 
 %{\color{blue} This section will ultimately be moved to the end of the paper or included as Supplemental Material.  Warning on notation. The following symbols are identical:  $$ u=v_1, \quad v=v_2, \quad w=v_3.$$ }
 
 We start with the governing equations \eqref{lin-momentum}-\eqref{stress-Cauchy-viscous}   and apply the scaling relations  \eqref{scaling1}-\eqref{nabla-bar}  that give the  dimensionless  operator $\bar\nabla$, the material constants  $\bar\alpha_i$, $\bar k_i$, the parameter  groups  $\mathcal R$, $\Er$ and $\mathcal A$, and the dimensionless fields $\bar\bv$ and $\bar p$.
%$$ \bar x=\frac{x}{L}, \, \bar y=\frac{y}{d}, \, \bar z=\frac{z}{L}; \quad \bar u=\frac{u}{V}, \, \bar v=\frac{v}{V}, \, \bar w=\frac{w}{V}, \bar t =\frac{t}{T}, \, T:=\frac{L}{V}. $$
%Let $\varepsilon=\frac{d}{L}$ denote the aspect ratio of the domain and use and the dimensionless groups as before, and keeping the notation $\bv, \mathbf N$ to denote the dimensionless fields, we have
%\begin{align*}\bar \nabla=&(\partderiv{}{\bar x}, \varepsilon^{-1} \partderiv{}{\bar y}, \partderiv{}{\bar z}),\\
%\\\bar\nabla_2=& (\partderiv{}{\bar x}, \varepsilon^{-2} \partderiv{}{\bar y}, \partderiv{}{\bar z}),\\\bar{\mathbf v}=& (\bar u, \bar v, \bar w).\end{align*}
%We let $\Er= \frac{\eta_0 LV}{K}$, where $\eta_0$ is a typical viscosity.
%The dimensionless equations are now in the form
\begin{eqnarray}
\Re \dot{\bar\bv}&=&\bar\nabla \cdot \sigma_N, \label{lin-momentum-scaled-new}\\
\bar\gamma_1 \dot{\mathbf{n}}&=&{\Er}^{-1}\big(\bar\nabla \cdot  \left(\frac{\partial  \wof}{\partial  \bar\nabla \mathbf{n}} \right)-\frac{\partial
 \wof}{\partial  \mathbf{n}}\big)
 %\nonumber\\&&
 +\bar\gamma_1 \boldsymbol{\Omega} -\bar\gamma_2 \mathbf{An}+\lambda_{\tiny{\text{L}}}\bn,\label{ang-momentum1}\\
\bar\nabla\cdot\bar\bv&=&0,\label{incompressibility-scaled-new}\\
\bn\cdot\bn&=&1, 
\end{eqnarray}
 The Oseen-Frank energy is as in \eqref{Oseen_Frank} with $k_i$ and $\nabla$ replaced with $\bar k_i$ and $\bar \nabla$
\begin{eqnarray}
 \sigma_N&=&-\bar p \mathbf{I} -\bar\nabla \mathbf{n}^T\frac{\partial \wof}{\partial \bar\nabla \mathbf{n}} 
+\hat{\sigma}_N+\sigma_a,\label{stress-Cauchy-total-scaled}\\
 \hat{\sigma}_N&=&\bar\alpha_1\left(\mathbf{n}\cdot \mathbf{A}\mathbf{n}\right)\mathbf{n}\otimes  \mathbf{n}+\bar \alpha_2\mathbf{N}\otimes\mathbf{n}+\bar\alpha_3\mathbf{n}\otimes \mathbf{N}+\bar\alpha_4 \mathbf{A}%\nonumber\\&&
+\bar\alpha_5 \mathbf{A} \mathbf{n}\otimes \mathbf{n}+\bar\alpha_6\mathbf{n}\otimes  \mathbf{A} \mathbf{n},\label{stress-Cauchy-viscous-scaled-new}\\
\sigma_a&=&-\mathcal A\,  \mathbf{n}\otimes \mathbf{n}, \label{stress-active-scaled-new}
\end{eqnarray}
where  $$2\mathbf{A} =\bar\nabla \bar{\mathbf{v}}+(\bar\nabla \bar{\mathbf{v}})^T,
\quad  2\boldsymbol{\Omega}=\bar\nabla \bar{\mathbf{v}} -(\bar\nabla \bar{\mathbf{v}})^T \mbox{ and }
\mathbf{N}=\dot{\mathbf{n}}-\boldsymbol{\Omega}  \mathbf{n}.$$
We recall that the parameter $\varepsilon=\frac{d}{L}$ in the definition of $\bar\nabla=(\partderiv{}{\bar x}, \varepsilon\partderiv{}{\bar y}, \partderiv{}{\bar z})$ denotes the aspect ratio of the nematic domain.

The symbols are used interchangeably but  represent the same fields :  $ u=v_1, \, v=v_2, \, w=v_3.$

We carry out the following calculations for the equal constant energy $K_1=K_2=K_3:=K, K_4=0.$ In this case,
\begin{align}
    \partderiv{W}{\bar\nabla \bn}=&K\left[\begin{matrix} 
        n_{1\rx} & \frac{1}{\varepsilon} n_{1\ry} & n_{1\rz}\\
        n_{2\rx} &\frac{1}{\varepsilon}{n_{2\ry}} &n_{2\rz}\\   
         n_{3\rx} &\frac{1}{\varepsilon}{n_{3\ry}} &n_{3\rz}\\ 
    \end{matrix}\right],\\
    \bigg(\bar\nabla\cdot  \partderiv{W}{\bar\nabla \bn}\bigg)_{i} =&K \big(\partderivtwo{n_i}{\rx} + \frac{1}{\varepsilon^2} \partderivtwo{n_i}{\ry} +\partderivtwo{n_i}{\rz}\big), \quad i=1,2,3. \label{scaled-Nforce-new}
\end{align}
Now we calculate the scaled nematic stress tensor:
%\begin{align}    \mathbf A=\left[\begin{matrix}\ru_\rx & \frac{1}{2}(\rv_{\rx}+\frac{1}{\varepsilon}\ru_{\ry} ) & \ru_{\rz}+\rw_{\rx}\\  \frac{1}{2}(\rv_{\rx}+\frac{1}{\varepsilon}\ru_{\ry}) &\frac{1}{\varepsilon}\rv_\ry& \frac{1}{2}(\rv_{\rx}+\frac{1}{\varepsilon}\rv_{\rz} )  \end{matrix} \right]\end{align}

\begin{align} {\label{A-Omega}}
\mathbf{\Omega}&= \left[\begin{matrix} 0& \frac{1}{2}(\frac{1}{\varepsilon}\partderiv{\ru}{\ry}-\partderiv{\rv}{\rx})& \frac{1}{2}(\partderiv{\ru}{\rz}-\partderiv{\rw}{\rx})\\ -\frac{1}{2}(\frac{1}{\varepsilon}\partderiv{\ru}{\ry}  -\partderiv{\rv}{\rx}) & 0& \frac{1}{2}(\partderiv{\rv}{\rz}-\frac{1}{\varepsilon}\partderiv{\rw}{\ry}) \\-\frac{1}{2}(\partderiv{\ru}{\rz}-\partderiv{\rw}{\rx}) &- \frac{1}{2}(\partderiv{\rv}{\rz}-\frac{1}{\varepsilon}\partderiv{\rw}{\ry})&0
\end{matrix} \right],\\
\mathbf{A}&= \left[\begin{matrix} \partderiv{\ru}{\rx}& \frac{1}{2}(\frac{1}{\varepsilon}\partderiv{\ru}{\ry}+\partderiv{\rv}{\rx})& \frac{1}{2}(\partderiv{\ru}{\rz}+\partderiv{\rw}{\rx})\\ \frac{1}{2}(\frac{1}{\varepsilon}\partderiv{\ru}{\ry} +\partderiv{\rv}{\rx}) & \frac{1}{\varepsilon}\partderiv{\rv}{\ry}& \frac{1}{2}(\partderiv{\rv}{\rz}+\frac{1}{\varepsilon}\partderiv{\rw}{\ry}) \\\frac{1}{2}(\partderiv{\ru}{\rz}+\partderiv{\rw}{\rx}) &\frac{1}{2}(\partderiv{\rv}{\rz}+\frac{1}{\varepsilon}\partderiv{\rw}{\ry})&\partderiv{\rw}{\rz}
\end{matrix} \right]
\end{align}
We now proceed to calculating the asymptotic limit of the equation of balance of linear momentum \eqref{lin-momentum-scaled-new}, upon setting $n_2=0$. In particular, we determine the order of magnitude of each term according to the powers of $\frac{1}{\varepsilon}$. Let us calculate 
\begin{equation}
    Q:= \bar\nabla\bn^T\partderiv{\wof}{\bar\nabla \bn}= K\left[\begin{matrix} n_{1,\bar x}^2+ n_{3,\bar x}^2 & 0 & n_{1,\bar x} n_{1,\bar z}+ n_{3,\bar x}n_{3,\bar z} \\
    0&0&0\\n_{1,\bar x} n_{1,\bar z}+ n_{3,\bar x}n_{3,\bar z}&0& n_{1,\bar z}^2+ n_{3,\bar z}^2 
    \end{matrix}\right]
\end{equation}
Now, calculate
\begin{equation}
    \bar\nabla\cdot Q\large|_i =\partderiv{Q_{i1}}{\bar x} +\partderiv{Q_{i3}}{\bar z}  = O(1), \quad i=1,2,3.
\end{equation}
(Note that terms involving $\frac{1}{\varepsilon}\partderiv{}{\bar y}$ do not appear in the previous calculation). Next, we calculate the leading terms of  equation \eqref{lin-momentum-scaled-new}, taking into account the calculations \eqref{A-Omega} above:

\begin{align}
    0=&\frac{1}{2\varepsilon^2}\big(\alpha_4+ n_1(\alpha_2+\alpha_5)\big)\partderivtwo{\bar u}{\bar y} +O(\frac{1}{\varepsilon})\label{balance-lin-momenttum-leading1}\\
    0=&O(\frac{1}{\varepsilon}),\\ 0=&\frac{1}{2\varepsilon^2}\big(\alpha_4+ n_3(\alpha_2+\alpha_5)\big)\partderivtwo{\bar w}{\bar y} +O(\frac{1}{\varepsilon})\label{balance-lin-momenttum-leading3}
\end{align}

\subsection{Rate of dissipation $\mathcal R$ for exponentially thin interfaces}\label{RateDissipationExp} 
\begin{figure}[htbp!] 
\begin{center}
\includegraphics[width=0.60\textwidth]{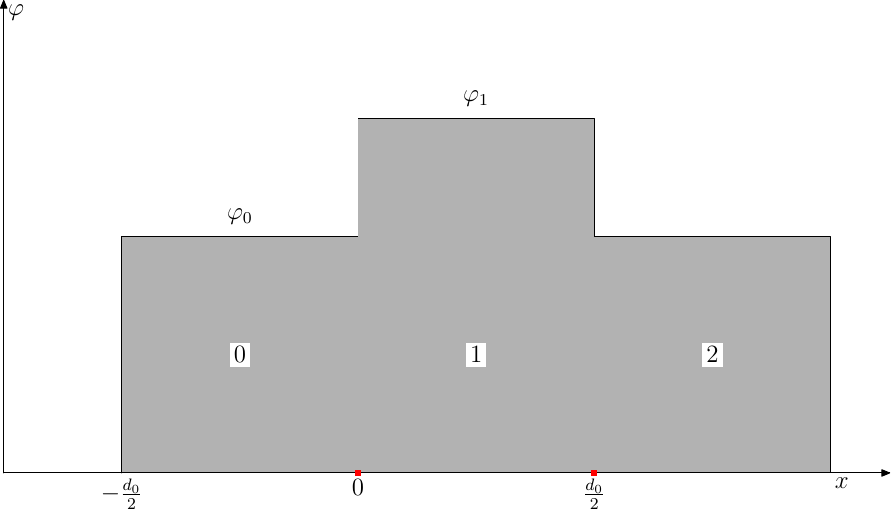}
\caption{This is the illustration of a chevron unit: (1) half domain of length $\lambda_0$ with $\varphi=\varphi_0$; (2)  half domain of length $\lambda_0$ with $\varphi=\varphi_1$; (3) interface $\varphi_0\to \varphi_1$ between  both half domains; (4) interface  $\varphi_1\to \varphi_0$ at the end of the second  half. (That is, the unit chevron consists of regions 0 and 1, and the two interfaces marked in red. For convenience, we mark region 2, a copy of region 0. }
\label{C-diagram}
\end{center}
\end{figure}

We consider the total rate of dissipation $\mathcal R$ as in \eqref{T-dissipation}. We evaluate it on approximations of  angular fields $\varphi$ as in \eqref{regular+internal}, with a single profile sketched in the diagram \ref{C-diagram}.
We will find that $\mathcal R(\lambda_0) $is monotonically increasing and, so, it fails to select a pattern wavelength by the principle of minimum dissipation.

To estimate $\mathcal  R$, we assume that the pattern is formed by  $n$ repeated chevron units located next to each other, occupying a total length $L$. Each individual chevron (figure \ref{C-diagram}) is represented by 
a field of the form \eqref{regular+internal}. This allows us to estimate the total rate of dissipation a 
\begin{equation}
\mathcal  R=\sum_{i=1}^n \mathcal R_i= n\mathcal R_0, \label{T-dissipation}
\end{equation}
where $\mathcal R_0$ represents the rate of dissipation of the representative unit chevron.  Let us estimate $\mathcal R_0$:
\begin{equation}
    \mathcal R_0= \int_{-\lambda_0}^{\lambda_0+\epsilon}\Delta(x)\,dx
\end{equation}
Note that the upper limit of the integral is meant to include the interface on the right end of the chevron.
Using \eqref{A-chevron}, we estimate 
\begin{align}
    \mathcal R_0=& \int_{-\lambda_0}^{0} \Delta(2\varphi_0)\,dx + \int_0^{\lambda_0} \Delta(2\varphi_1+2I\Phi(\frac{x}{\epsilon})\,dx+ \int_{\lambda_0}^{2\lambda_0} [\Delta(2\varphi_0+2I\Phi(\frac{x-{\lambda_0}}{\epsilon})-\Delta(2\varphi_0)] \, dx +O(\epsilon),
    \end{align}
 where the last integral only involves the contribution of the interface.  With this, we can write \eqref{T-dissipation}
 as 
 \begin{align}
     \mathcal R= &n\Delta(2\varphi_0) {\lambda_0}+ n \int_0^{\lambda_0} \Delta(2\varphi_1+2I\Phi)\,dx+n\int_{\lambda_0}^{2\lambda_0} \Delta(2\varphi_0+2I\Phi)\,dx- n\Delta(2\varphi_0) {\lambda_0}\\
     =&  \frac{L}{2\lambda_0} [\int_0^{\lambda_0} \Delta(2\varphi_1+2I\Phi)\,dx+\int_{\lambda_0}^{2\lambda_0} \Delta(2\varphi_0+2I\Phi)\,dx]
 \end{align}
The quantity  $\mathcal R= \mathcal R(\lambda_0)$ is a function of $\lambda_0$; to find its minimum, we calculate
\begin{align}
  \frac{1}{2}  \frac{d\mathcal R}{d \lambda_0}=&-\frac{L}{4\lambda_0^2}\int_{0}^{\lambda_0}\Delta_N(2\varphi_1+ 2 I\Phi(\bar x, 0)))\,dx +  \frac{L}{4\lambda_0}\Delta_N(2\varphi_1+ 2 I\Phi({\lambda_0}, 0))\nonumber \\
    -&\frac{L}{4\lambda_0^2}\int_{\lambda_0}^{2\lambda_0}\Delta_N(2\varphi_0+ 2 I\Phi(\bar x, 0)))\,dx
   +  \frac{L}{2\lambda_0}\Delta_N(2\varphi_0+ 2 I\Phi(\lambda_0, 0))-  \frac{L}{4\lambda_0}\Delta_N(2\varphi_0+ 2 I\Phi({\lambda_0}, 0)).
\end{align}
Note that 
\begin{align}
\Delta_N(2(\varphi_1+ I\Phi(\bar x,0)) = &
\Delta_N(2\varphi_1)+ \partderiv{\Delta_N}{\varphi}(2\varphi_1)I\Phi(\bar  x, 0)+ O(\epsilon)\nonumber\\= & \Delta_N(2\varphi_1)+ \partderiv{\Delta_N}{\varphi}(2\varphi_1)(\varphi_0-\varphi_1)e^{-\omega_1\frac{x}{\epsilon}}+ O(\epsilon), \nonumber\\
\int_{0}^{\lambda_0}\Delta_N(2\varphi_1+ 2 I\Phi(\bar x, 0))\,dx =& {\lambda_0}\Delta_N(2\varphi_1)+ \partderiv{\Delta_N}{\varphi}(2\varphi_1)\frac{\epsilon}{\omega_1}(\varphi_0-\varphi_1)(1- e^{-\omega_1\frac{\lambda_0}{\epsilon}})+ O(\epsilon).
\end{align}
Likewise,
\begin{align}
\Delta_N(2(\varphi_0+ I\Phi(\bar x,0)) = &
\Delta_N(2\varphi_0)+ \partderiv{\Delta_N}{\varphi}(2\varphi_0)I\Phi(\bar  x, 0)+ O(\epsilon)\nonumber\\= & \Delta_N(2\varphi_0)+ \partderiv{\Delta_N}{\varphi}(2\varphi_0)(\varphi_1-\varphi_0)e^{-\omega_1\frac{x}{\epsilon}}+ O(\epsilon), \nonumber\\
\int_{\lambda_0}^{2\lambda_0}\Delta_N(2\varphi_0+ 2 I\Phi(\bar x, 0))\,dx =& {\lambda_0}\Delta_N(2\varphi_0)+ \partderiv{\Delta_N}{\varphi}(2\varphi_0)\frac{\epsilon}{\omega_1}(\varphi_1-\varphi_0)(e^{-\omega_1\frac{\lambda_0}{\epsilon}}- e^{-\omega_1\frac{2\lambda_0}{\epsilon}})+ O(\epsilon).
\end{align}
Hence
\begin{eqnarray*}
%\begin{align*}
&& \frac{1}{2}\frac{d\mathcal R}{d \lambda_0}= -\frac{L}{\lambda_0^2}\big[{\lambda_0}\Delta_N(2\varphi_1)+ \partderiv{\Delta_N}{\varphi}(2\varphi_1)\frac{\epsilon}{\omega_1}(\varphi_0-\varphi_1)(1-e^{-\omega_1\frac{\lambda_0}{\epsilon}}) \big]
-\frac{L}{4\lambda_0^2}\big[ {\lambda_0}\Delta_N(2\varphi_0)+ \partderiv{\Delta_N}{\varphi}(2\varphi_0)\frac{\epsilon}{\omega_1}(\varphi_1-\varphi_0)(e^{-\omega_1\frac{\lambda_0}{\epsilon}}- e^{-\omega_1\frac{2\lambda_0}{\epsilon}})\big]
 \\&&\quad 
 + \frac{L}{4\lambda_0}\big[ \Delta_N(2\varphi_1)+ \partderiv{\Delta_N}{\varphi}(2\varphi_1)(\varphi_0-\varphi_1)e^{-\omega_1\frac{\lambda_0}{\epsilon}} \big ]\\
 &&\quad
{ +\frac{L}{2\lambda_0}\big[\Delta_N(2\varphi_0)+ \partderiv{\Delta_N}{\varphi}(2\varphi_0)(\varphi_1-\varphi_0)e^{-\omega_1\frac{2\lambda_0}{\epsilon}} \big]}\\
 &&\quad{-\frac{L}{2\lambda_0}\big[ \Delta_N(2\varphi_0)+ \partderiv{\Delta_N}{\varphi}(2\varphi_0)(\varphi_1-\varphi_0)e^{-\omega_1\frac{\lambda_0}{2\epsilon}} \big]
 }\\
 &&\quad=  -\frac{L}{4\lambda_0^2}\big[ \partderiv{\Delta_N}{\varphi}(2\varphi_1)\frac{\epsilon}{\omega_1}(\varphi_0-\varphi_1)(1-e^{-\omega_1\frac{\lambda_0}{\epsilon}}) \big]+ \frac{L}{4\lambda_0}\big[  \partderiv{\Delta_N}{\varphi}(2\varphi_1)(\varphi_0-\varphi_1)e^{-\omega_1\frac{\lambda_0}{\epsilon}} \big ]\\
&& \quad -\frac{L}{\lambda_0^2}\big[\partderiv{\Delta_N}{\varphi}(2\varphi_0)\frac{\epsilon}{\omega_1}(\varphi_1-\varphi_0)(e^{-\omega_1\frac{\lambda_0}{\epsilon}}- e^{-\omega_1\frac{2\lambda_0}{\epsilon}})\big]+\frac{L}{\lambda_0}\big[ \partderiv{\Delta_N}{\varphi}(2\varphi_0)(\varphi_1-\varphi_0)e^{-\omega_1\frac{\lambda_0}{\epsilon}} \big]\\
&&\quad -\frac{L}{2\lambda_0}\big[\partderiv{\Delta_N}{\varphi}(2\varphi_0)(\varphi_1-\varphi_0)e^{-\omega_1\frac{\lambda_0}{\epsilon}} \big]
%\end{align*}   
\end{eqnarray*}
Denote
\begin{equation*}
    V(\lambda_0):= \frac{32\lambda_0^2}{\A^2 L}  \frac{d\mathcal R}{d \lambda_0}.
\end{equation*}
The following calculations are needed in order to determine the critical points of $\mathcal R$:
\begin{align*}
    \cos{2\varphi_0}=&\cos{2\varphi_1}= \frac{\gamma_1}{\gamma-2}, \quad 
    \sin^2{2\varphi_1}=\sin^2{2\varphi_0}= 1- (\frac{\gamma_1}{\gamma_2})^2,\\
    R(2\varphi_1)=&R(2\varphi_0) =\bar\alpha_1(1-\frac{\gamma_1^2}{\gamma_2^2}) -(\bar\alpha_2+\bar\alpha_3-\bar\alpha_5+\bar\alpha_6)\frac{\gamma_1}{\gamma_2} + \bar\alpha_3-\bar\alpha_2+ 2\bar\alpha_4 +\bar\alpha_5+\bar\alpha_6, \\
    \frac{dR}{d\varphi}(2\varphi_0)=&\frac{dR}{d\varphi}(2\varphi_1)= 2\alpha_1\sin{4\varphi_0}+ 2(\alpha_2+\alpha_3-\alpha_5+\alpha_6)\sin{2\varphi_0}\\
    g(2\varphi_0)=& \frac{1}{2}\big[\bar\alpha_1\big(\sin^2{2\varphi_0}-\sin{2\varphi_0}(1-\cos{2\varphi_0})\big)+ (\bar\alpha_2+\bar\alpha_5)(1-\cos{2\varphi_0})+ (\bar\alpha_6-\bar\alpha_5)(1+\cos{2\varphi_0})\big]\\   g(2\varphi_1)= & 
\end{align*}
Next, we calculate
\begin{align*}
\Delta_N(2\varphi_0)=&\frac{\A^2}{8}\sin^2{2\varphi_0}\frac{R(2\varphi_0)}{g^2(2\varphi_0)} \\
\Delta_N(2\varphi_1)=&\frac{\A^2}{8}\sin^2{2\varphi_1}\frac{R(2\varphi_1)}{g^2(2\varphi_1)}=\frac{\A^2}{8}\sin^2{2\varphi_0}\frac{R(2\varphi_0)}{g^2(2\varphi_1)}\\
\partderiv{\Delta_N}{\varphi}(2\varphi)=&\frac{\A^2}{8g^2(2\varphi)}\big(2\sin{4\varphi}R+\sin^2{2\varphi}(\frac{dR}{d\varphi}-2\frac{R}{g(2\varphi)})\big)\\
%\partderiv{\Delta_N}{\varphi}(2\varphi_1)=&
\end{align*}
With the above calculations, we arrive at the estimate
\begin{equation}
  \frac{1}{2}  \frac{d\mathcal R}{d \lambda_0}= {e^{-\omega_1\frac{2\lambda_0}{\epsilon}}}\big( e^{\omega_1\frac{\lambda_0}{\epsilon}}-1\big)>0.
\end{equation}
Hence, the exponential thickness of the transition regions is not able to single out a length scale of the domains.  This is due to the special scaling of the governing equation, with no decaying terms in $x$. This prevents a competition between the rate of energy dissipation of the bulk domains with respect to that of the transition regions.

\begin{figure}
%\begin{center}
 \begin{subfigure}[b]{0.6\textwidth}
\includegraphics[width=7cm]{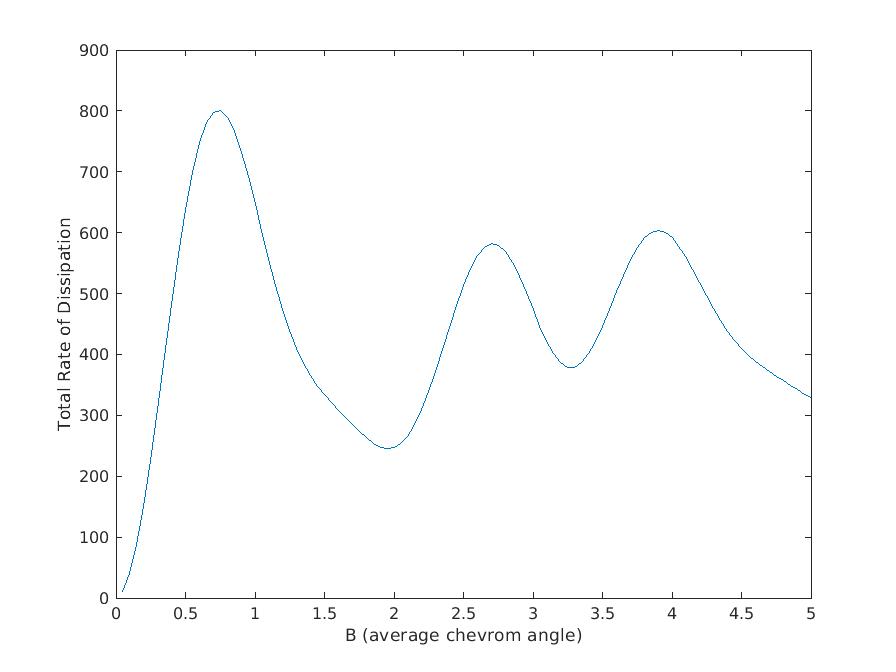} 
%\caption{Potential}
 \end{subfigure}
 \hfill 
\begin{subfigure}[b]{0.6\textwidth}
\includegraphics[width=7cm]{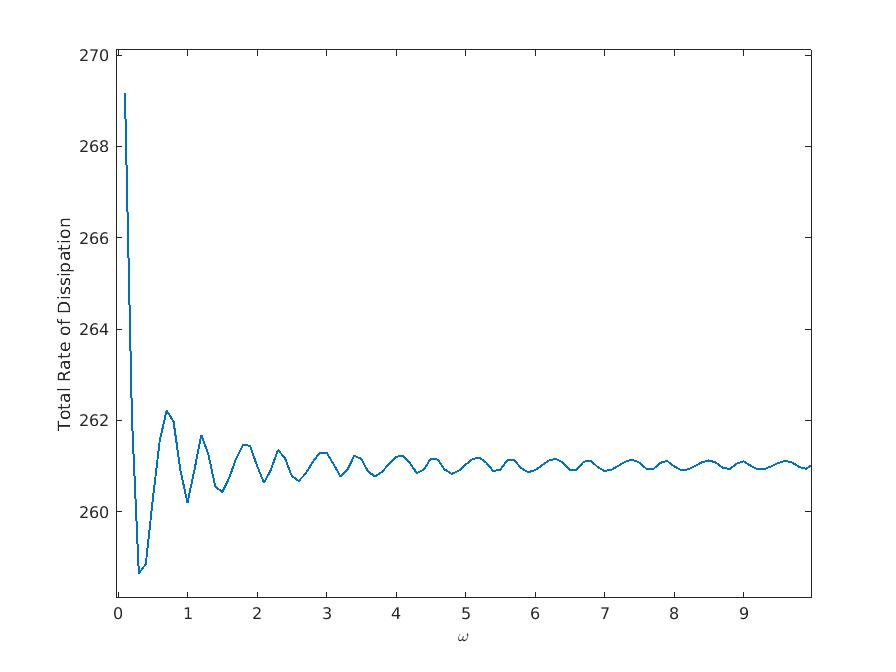}
%\caption{Directional angle $\varphi(x)$ }
 \end{subfigure}
 \caption{(a) Total rate of dissipation with respect to the averaged angular amplitude $B$, for $\omega$ fixed at a value in the experimental range, $\omega=.044$.  (b) Total rate of dissipation with respect to the wave-length $\omega$ upon fixing the chevron angle to $\varphi=1.055$} 
% \label{Potential-fig}
 %\end{center}
 \end{figure}

\subsection {Domains separated by line defects}
We assume that the pattern consists of alternating regions of angular values $\varphi_0$, $\varphi_1$ separated by line defects.  Following the analogous approach to calculating the liquid crystal energy associated with a line (or plane) defect as in the Rapini-Papoular theory \cite{de1993physics}, we now assume that energy dissipation is also associated, in this case,  to velocity jumps and it has the analogous form, 
\begin{equation}
 \Delta_{\text{jump}}= C [V], \label{Delta-jump}
\end{equation}
with $C>0$ having dimensions of rate of  energy dissipation. 
  The problem also becomes qualitatively analogous to that of martensite to austenite  phase transformation, where the competition between the  interfacial and bulk energies  provides the length scale of the pattern. In this section, we follow the inverse approach of estimating the dissipation parameter $C$  corresponding to the line separation $\lambda_0$ and the interface thickness $\delta$  given in \eqref{delta-asymptotic}. 
  
  Let us consider a pattern of $2n>0$ bands (n full chevrons)  each of angle $\varphi_0$ and $\varphi_1$, respectively, separated by $2n-1$  interfaces. As in section\ref{internal-layer}  $\lambda_0>0 $ denotes the line spacing with and $\delta$,  the  thickness of the interface between between neighboring regions $\varphi_0, \varphi_1$.
  Denoting
  \begin{equation}
      \Delta_1=\Delta_N(2\varphi_1), \,\, \Delta_0= \Delta_N(2\varphi_0), \,\, \Delta_{01}=\frac{1}{2}(\Delta_0+\Delta_1),
  \end{equation}
  the total rate of energy dissipation of the pattern is set as 
  \begin{equation}
      \mathcal R(\lambda_0)= 2n\Delta_{01}\lambda_0 + (2n-1)C[V]\delta. \label{TotalRateSharpI}
  \end{equation}
The minimization is done under the constraint
\begin{equation}
    2n\lambda_0+(2n-1)\delta=L.
\end{equation}
  Substituting the latter into  $\mathcal R$ in \eqref{TotalRateSharpI} gives
  \begin{equation}
      \mathcal R(\lambda_0)=\Delta_{01}\frac{(L+\delta)\lambda_0}{\lambda_0+\delta}+ (\frac{L+\delta}{\lambda_0+\delta}-1)C[V]\delta. \quad %\delta.%=\Gamma\lambda_0.
  \end{equation}
% Equivalently,
 % \begin{equation}  \mathcal R(\lambda_0)=\Delta_{01}\frac{(L+\Gamma\lambda_0)}{1+\Gamma}+ \big(\frac{L+\Gamma\lambda_0}{\lambda_0(1+\Gamma)}-1\big)C[\varphi]\Gamma\lambda_0. \end{equation}
  Note that the function $
  \mathcal R(\lambda_0)>0.$ Let us calculate
\begin{equation}
\mathcal R'(\lambda_0)= \frac{\Delta_{01}}{(\lambda_0+\delta)^2}\big((L-1)\lambda_0+ (L+\delta)\delta\big)-\frac{L+\delta}{(\lambda_0+\delta)^2} C [V]\delta, \label{dissipation-derivative}
\end{equation}
It is easy to check that 
\begin{equation}
    \mathcal R'(\lambda_0)=0 \quad \text {holds, provided} \quad C=\frac{\Delta_{01}}{[V]} 
\end{equation}
The latter step completes the determination of the constitutive equation for the energy dissipated in the chevron transition interfaces with respect to that of the change in angle. Furthermore, we arrive at the expression
\begin{equation}
    \Delta_{\text{jump}}=\frac{1}{2}(\Delta_0+\Delta_1).
\end{equation}
This indicates that the energy dissipated on the interfaces is of the order of magnitude of that lost in the constant angle region. 

\section{Numerical solution of the scaled equation}
 We  now present a simulation of the scaled equation \eqref{angle-pattern-base} as shown in figure 10.  The variables and field shown next are dimensionless.
 
 \begin{figure}[hbt!]
\begin{center}
 \begin{subfigure}[t]{.495\textwidth}
 \begin{center}
  \includegraphics[width=1\textwidth]{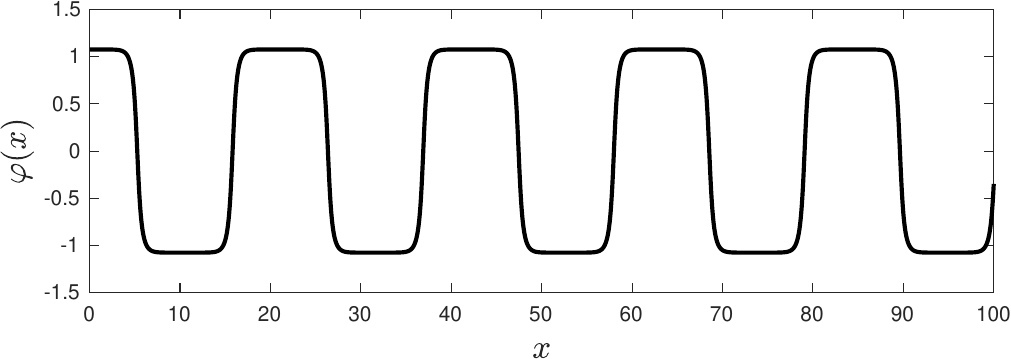} 
  \caption{Profile of the angle of director alignment}
  \bigskip
   \end{center}
 \end{subfigure}
 \begin{subfigure}[t]{.495\textwidth}
 \begin{center}
\includegraphics[width=0.95\textwidth]{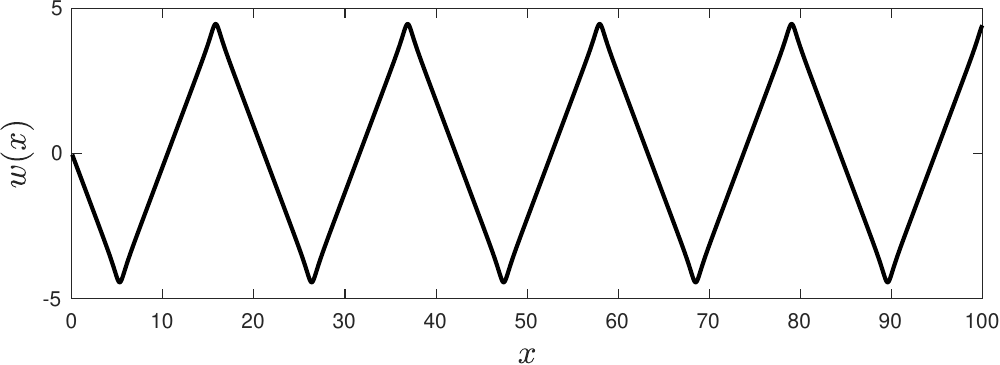} 
  \caption{Profile of the velocity field}
   \end{center}
 \end{subfigure}
 \caption{Numerical simulations of the  velocity field (b)  and directional angle (a) obtained as solutions of  equations \eqref{velocity_gradient-eqn-reduced} and \eqref{angle-pattern-base}, respectively. }
 %The spatial variable is the scaled as $\frac{x}{\epsilon}$, where $\epsilon$ is defined in \eqref{def_espilon}, and {$w(x)$ has unitis of $\frac{\mu m}{\text{sec}}$ } and $\varphi$ in radians.    } 
\label{fig_scaledEqn}
\end{center} 
\end{figure}\begin{figure}[hbt!] \label{wave-lenth-estimates}
 \begin{subfigure}[t]{.45\textwidth}
    \centering
    \includegraphics[width=0.95\linewidth]{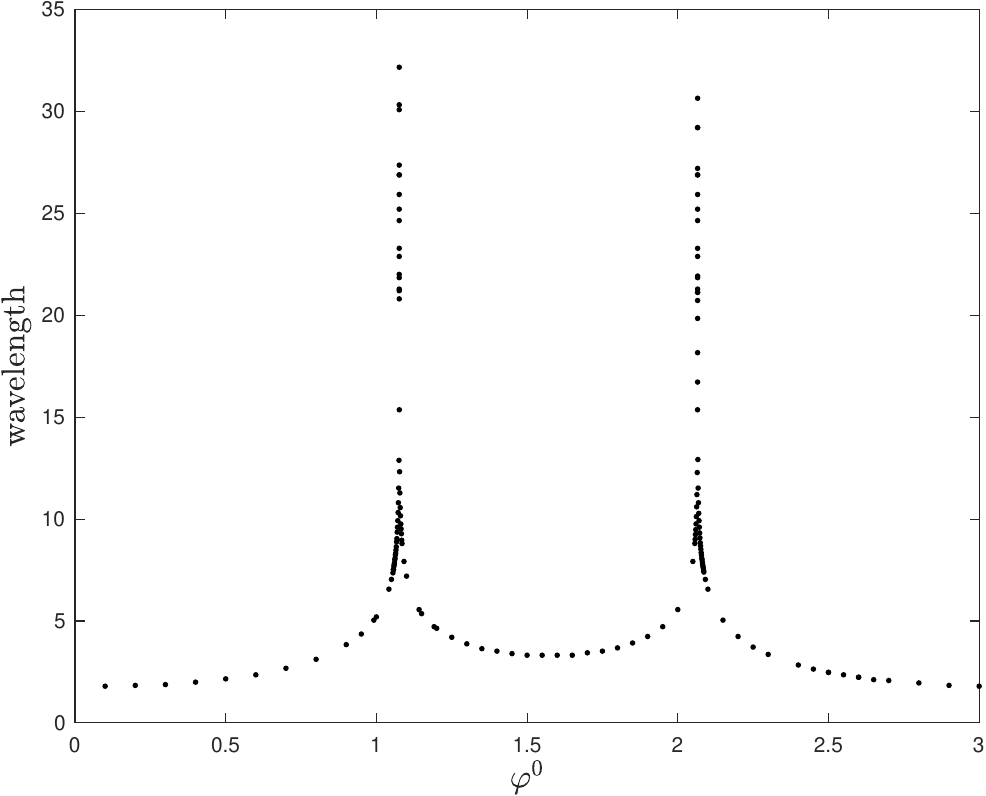}
    \caption{Wavelength of $\varphi$ respect to the initial value $\varphi^0$}
    \label{fig:wavelengthVSinit}
\end{subfigure}
 \begin{subfigure}[t]{.45\textwidth}
    \centering
    \includegraphics[width=0.95\linewidth]{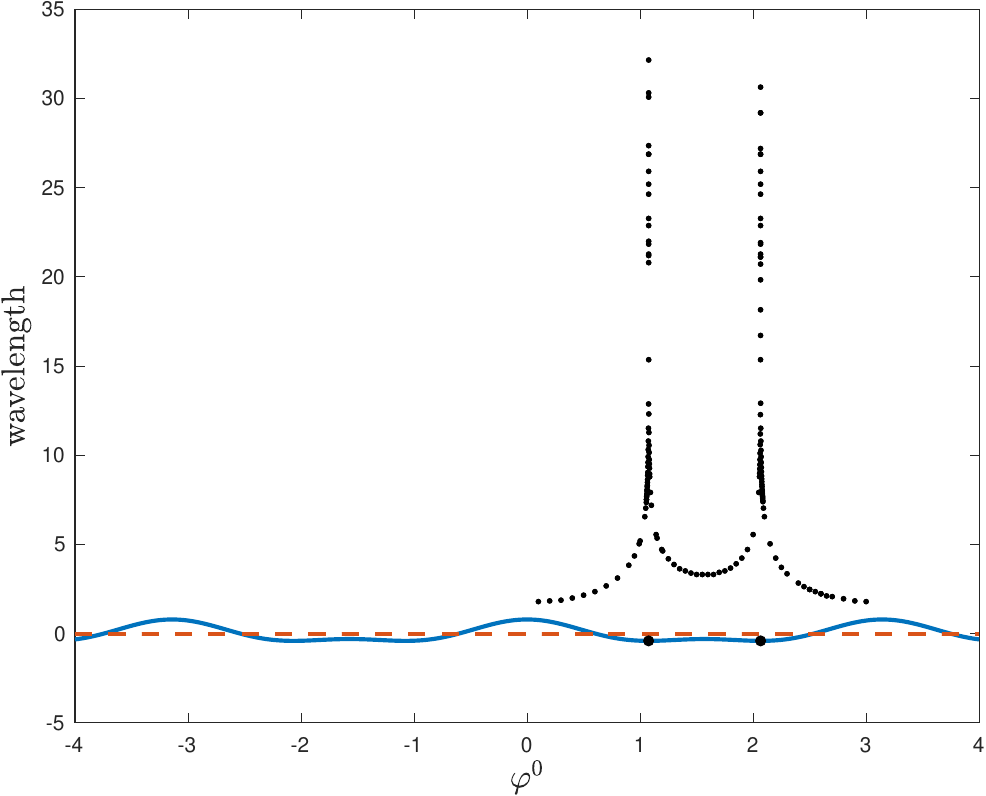}
    \caption{Superimposed wavelength plot  (\ref{fig:wavelengthVSinit}) on the potential plot of figure \ref{fig:potential}}
    \label{fig:wavelengthONpotential}
    \end{subfigure}
    \caption{Wave length comparisons}
\end{figure}
  \begin{equation*}\frac {d^2\varphi}{d\bar x^2}+(-\frac{\bar\gamma_1}{\bar\gamma_2}\sin(2\varphi)+\frac{1}{2}\sin(4\varphi))=0.  
  \end{equation*}
 % Then figure \ref{fig_scaledEqn} show the numerical results. 
Figure \ref{fig:wavelengthVSinit}
and \ref{fig:wavelengthONpotential}
 are plots of the chevron wavelength with differential initial values $\varphi^0$ when we solve the scaled equations \eqref{angle-pattern-base} as an initial value problem $\varphi(0)=\varphi^0$ and $\frac{d\varphi}{dx}(0)=0$.
 These plots show that as the initial values closes to the critical values $\varphi_0$ and $\varphi_1$ in \eqref{equilibrium-nonzero},   the wavelength increases. This is consistent with results indicated by the phase portrait \ref{fig:phaseportrait}.  The solutions are the red heteroclinic orbits. Otherwise, solutions are indicated as either black or purple periodic orbits.  Especially, figure \ref{fig:wavelengthONpotential} are the superimpose figure \ref{fig:wavelengthVSinit} with figure \ref{fig:potential}. It is easier to show the relation of the angles to the wavelength. 

 Note that, in figure \ref{fig_scaledEqn} (a), The direction angle $\varphi$ is computed with the scaled variable $\bar{x}=\frac{x}{\epsilon}$ defined in equation \eqref{def_espilon}. 
 However, the velocity $w(x)$ in   figure \ref{fig_scaledEqn}(b)  is computed by substitute $\varphi(\bar{x})$  into the original variable with  equation \eqref{velocity_gradient-eqn-reduced}, i.e.,
  \begin{equation}
    g(2\varphi(\bar{x}))  \partderiv{w}{x} = \mathcal{A}\sin(2\varphi(\bar{x})),\label{eqn:velocity_mixed_notation}
  \end{equation}
  To make the results consistent, the $w(\bar{x})$ should be computed. However,  the velocity in the  rescaled variable $w(\bar{x})$ is the result showed in figure \ref{fig_scaledEqn}(b) multiplied by a factor $\epsilon$.

Finally, we include with a plot of the $\frac{1}{4}$ of the heteroclinc orbit that joints the saddle points $\varphi_0$ and $\varphi_1$.
\begin{figure}
\centering
\includegraphics[height=2.8cm]{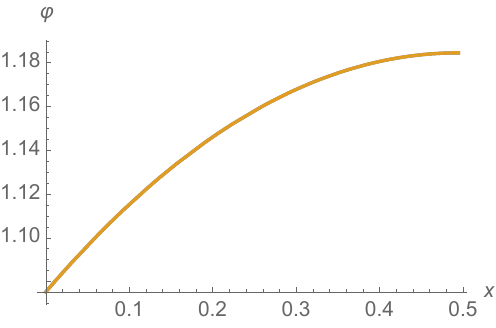}
\caption{Profile of half of the heteroclinic orbit of the phase plane that   connects   $\varphi_0$ and $\varphi_1$, in the  dimensionless  scale $ x$
  }
\end{figure}

\end{document}